\documentclass[aps,preprint]{revtex4}%
\usepackage{graphicx}
\usepackage{amssymb}
\usepackage{graphicx}
\usepackage{amsmath}
\usepackage{amsfonts}
\usepackage{color}%
\providecommand{\U}[1]{\protect\rule{.1in}{.1in}}
\providecommand{\U}[1]{\protect\rule{.1in}{.1in}}

\usepackage{graphicx}
\usepackage{mathtools}

\begin{document}

\title{Non-equilibrium phenomena in superconductors probed by femtosecond
time-domain spectroscopy}
\author{J. Demsar}
\affiliation{Institute of Physics, Johannes Gutenberg-University Mainz, 51099 Mainz, Germany}

\begin{abstract}
Development of ultrafast lasers and non-linear optical techniques over the
last two decades provides tools to access real-time dynamics of low energy
excitations in superconductors. For example, time-resolved THz spectroscopy
and time- and angular-resolved photoemission spectroscopy provide access to
the real-time dynamics of the superconducting gap amplitude. Such studies
enable determination of microscopic parameters like quasi-particle
recombination rates, pair-breaking rates and electron-boson coupling
constants. Recently, intense THz pulses have been used to probe the
non-linear dynamics, including observation of collective modes. Moreover,
using low frequency electromagnetic pulses, there are several reports of
amplification of superconductivity in both conventional and unconventional
superconductors. Starting with a brief historical overview of the pioneering
work, where non-equilibrium phenomena in superconductors were investigated
using quasi-continuous excitation, we review some of the insights that are
provided by using real-time approaches. We focus on conventional BCS
superconductors, whose ground state is reasonably well understood, and
address similarities and open questions related to the corresponding studies
in high-T$_{c}$ superconductors.

\end{abstract}

\maketitle

\section{Introduction}

\label{intro}

The field of non-equilibrium superconductivity \cite{Tinkham,Gray} started
soon after the first observation of the superconducting gap using optical
spectroscopy \cite{TinkhamGlover} and the development of the BCS theory.
Given that small energy gap $\Delta$ on the order of a meV (depending on the
material), superconductors were first considered as potential detectors for
far-infrared radiation \cite{Burstein}. For such applications, understanding
the dynamical processes in a superconductor as a function of base
temperature, photon energy and excitation density is crucial. On the other
hand, even more fascinating observations followed in early 1960's, where
superconductivity in Al and Sn was enhanced (increasing the critical
current, the superconducting gap and even the critical temperature) under
illumination with electromagnetic radiation at sub-gap frequencies \cite%
{Wyatt,Dayem,Klapwijk,Clarke}. Most of the early studies were performed in a
quasi-continuous excitation regime, with the characteristic timescales of
the processes only estimated. In the last two decades or so, however, the
time-resolved techniques, with their ability to probe the evolution of
optical constants, electronic band structure, magnetization and crystal
structure with femtosecond time resolution contributed to further our
understanding of out-of-equilibrium superconductors.

As noted, non-equilibrium superconductivity has been an intriguing field of
the modern solid state physics since the 1960's \cite{Tinkham}. Here, we
first briefly review the results of the fascinating studies performed in
1960's and 1970's - for more details we refer the reader to Refs. \cite%
{Tinkham,Gray}. Then we turn to studies of real-time dynamics, driven by the
rapid development of femtosecond lasers and the related non-linear optical
methods. While the first real-time experiments with picosecond time
resolution were performed on conventional superconductor Pb \cite{Federici},
the field started to blossom following the early studies on cuprate
superconductors \cite%
{Han,Chwalek,Albrecht,Stevens,DemsarYBCO,kabanov,Kaindl,THzAveritt,Segre}.
The first systematic real-time studies of dynamics in conventional
superconductors were performed substantially later \cite%
{Carr,MgB2,Lobo,Beck1}. Such studies still represent a small fraction of the
body of work performed on superconductors, still largely dominated by
studies of cuprate and pnictide high temperature superconductors.
Nevertheless, we focus on the results obtained on standard s-wave
superconductors and address the differences and peculiarities of the results
obtained on unconventional high temperature superconductors after that. For
a review, focusing on ultrafast spectroscopy in correlated high temperature
superconductors, we refer the interested reader to a recent review \cite%
{Giannetti}.

\subsection{Dynamics following the above gap excitation of a superconductor}

\label{DynHistory}

We should first note that, in general, there are two types of
non-equilibrium that can be realized in superconductors \cite{Tinkham},
often referred to as the energy-mode and charge-mode imbalance. The two
types are related to the nature of excitations in superconductors, which
continuously changes from a hole-like (far below the Fermi level) to an
electron-like (far above the Fermi level) within the energy range of $%
\approx2\Delta$ around the Fermi level.

The first type of nonequilibrium is realized by perturation with charged
particles, e.g. by quasiparticle injection through a superconductor-metal
junction, resulting in net quasiparticle charge density (note the fractional
charge of quasiparticles close to $\Delta$). In this case, the populations
of the electron-like and the hole-like quasiparticle branches do not
coincide, and such a state is referred to as charge imbalance (or
branch-imbalance). Correspondingly, there is a difference in the chemical
potentials of quasiparticles and the condensate (see Figure 11.1 of Ref. 
\cite{Tinkham}).

The energy-mode imbalance is realized by exciting a superconductor with
charge-neutral drive, like electromagnetic radiation or ultrasound, or with
symmetrical tunnel injection in a superconductor-insulator-superconductor
junction. Here the nonequilibrium state is characterized by a
(non-equilibrium) distribution function, which is particle-hole symmetric;
i.e. the populations of the electron-like and the hole-like branches of the
quasiparticle spectrum are the same (see Figure 11.1 of Ref. \cite{Tinkham}%
). Often, this state can be approximated by an effective quasiparticle
temperature T$^{\ast}$. It is the energy-mode imbalance that is considered
in the case of photoexcitation discussed in this work.

The first non-equilibrium studies/effects were observed using injection of
quasiparticles through tunnel barriers \cite{Ginsberg62}, or by
(quasi-)continuous laser irradiation \cite{Parker72}. Here, photo-excitation
with light, whose frequency is larger than $2\Delta$, breaks Cooper pairs
and gives rise to excess quasiparticle (QP) density.

Experiments in multi-tunnel-junction configuration provided first estimates
of lifetimes of excited QPs. While radiative recombination lifetime was
theoretically estimated for Pb to $\approx0.4$ s at 2 K \cite{Burstein},
tunnel-junction experiments provided the upper limit for the average QP
lifetime in Pb to be of the order of $0.2$ $\mu$s at 1.44 K \cite{Ginsberg62}%
. It was argued that the relaxation proceeds via QP recombination, where QP
pairs recombine to form Cooper pairs by the emission of phonons with $%
\hbar\omega>2\Delta$ \cite{Ginsberg62,SchriefferGinsberg}. We note again,
that these experiments were performed under the constant drive, with the
relaxation time estimates based on a series of underlying assumptions \cite%
{Ginsberg62,SchriefferGinsberg}. One of the first systematic temperature
dependent studies using QP injection were performed in Al \cite{MillerDayem}%
. The authors noted that the errors of the extracted relaxation times are
relative only (due to the fact that the extracted absolute values of
relaxation time depend on the choice/uncertainty of the model parameters
used) \cite{MillerDayem,Gray69}. Despite the uncertainty in the absolute
values, the results revealed that at temperatures far below T$_{c}$ the
recombination rate followed exponential temperature dependence with $%
\tau_{R}^{-1}\propto \exp(-0.3\Delta/k_{B}T)$ \cite{MillerDayem}. This
behavior (apart from the factor 0.3 in the exponent) could be accounted for
by considering that the relaxation of the QP density, $n_{QP}$, reflects the
bi-particle recombination process (see Figure 1a).

\begin{figure}[ptb]
\includegraphics[width=120mm]{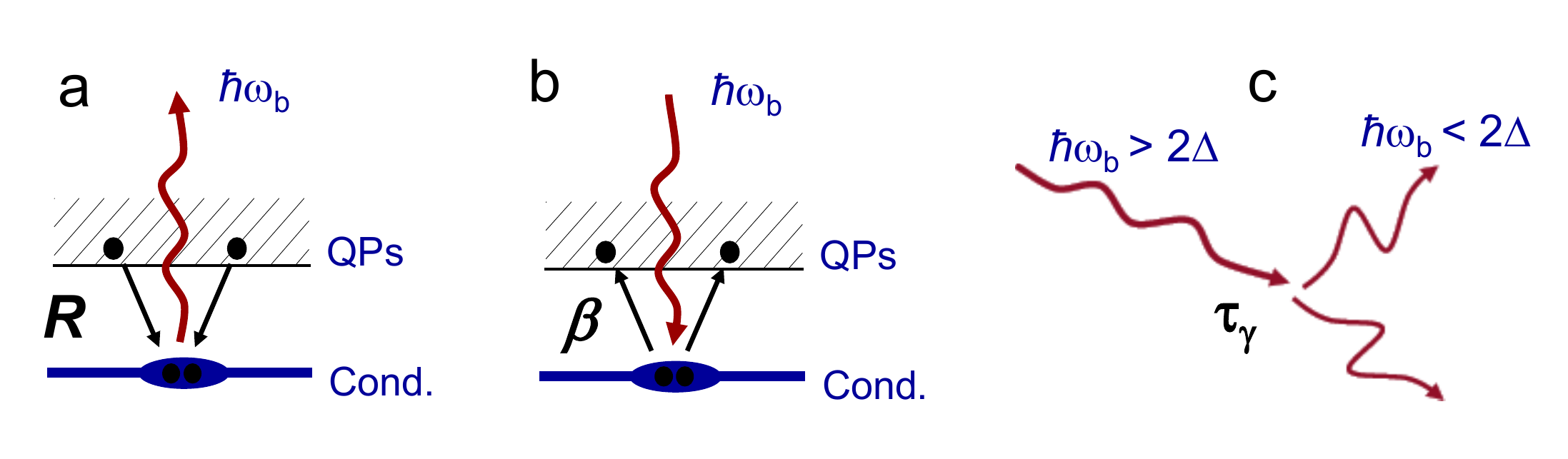} 
\caption{Different processes governing the relaxation dynamics of
photo-excited QPs in a superconductor: (a) recombination of two QPs into the
condensate via emission of 2$\Delta$ boson (phonon), (b) pair-breaking by
absorption of a boson, (c) decay of high energy bosons. The latter can
proceed via decay of bosons into the (dielectric) substrate of via
anharmonic decay into low-energy bosons, with energy less than 2$\Delta$.}
\label{Fig1}
\end{figure}

In this case, the dynamics of excess QP's is given by the bi-molecular rate
equation

\begin{equation}
\frac{dn_{QP}}{dt}=-Rn_{QP}^{2}  \label{bimolecular}
\end{equation}
where the recombination rate R is a microscopic parameter. These studies
were performed in the limit of weak perturbations, where the density of
injected (excess) QPs, $n_{QP}^{\prime}$, is small compared to the density
of thermally excited QPs, $n_{QP}^{T}$. With $n_{QP}=n_{QP}^{T}+n_{QP}^{%
\prime}$ and $n_{QP}^{\prime}\ll n_{QP}^{T}$, the above rate equation can be
approximated by $\frac{dn_{QP}^{\prime}}{dt}=-2Rn_{QP}^{T}n_{QP}^{\prime}$.
It is thus clear, that the exponential temperature dependence of relaxation
time can be attributed to the temperature dependence of the thermally
excited QPs, $n_{QP}^{T}\propto\sqrt{2\pi\Delta k_{B}T}\exp(-\Delta/k_{B}T)$.

We should note that most of these initial studies assumed that the
re-absorption of $\hbar\omega>2\Delta$ phonons (Figure 1b) emitted during
the QP recombination may be neglected. If these high energy phonons rapidly
decay into the substrate, or via anharmonic phonon decay (Figure 1c), it
will be the bi-particle recombination process that will be governing the
recovery of the SC state. It was Rothwarf and Taylor \cite{RothwarfTaylor},
that pointed out that in a general, phonon re-absorption cannot be neglected
(including the cases of high excitation densities). The slow decay of the
high frequency phonon populations leads in this case to the so-called phonon
bottleneck \cite{RothwarfTaylor}.

One of the first real-time studies of suppression of superconductivity with
light pulses was performed by Testardi \cite{Testardi}. He studied changes
in resistivity in optically thin lead films upon excitation with $\mu$s
pulses. He demonstrated, that the absorbed energy density required to
suppress superconductivity is substantially lower than the thermal energy,
required to heat up the superconductor up to T$_{c}$. To account for the
observation, Owen and Scalapino proposed the so-called $\mu^{\ast}$model 
\cite{OwenScalapino}, where under continuous excitation the QPs, the
condensate and the lattice are in thermal equilibrium, yet the chemical
potentials of QPs and the condensate are different. This assumption was
motivated by considering the thermalization times being faster than the QP
recombination times. According to the $\mu^{\ast}$ model, upon increasing
the excitation density, the superconductor should undergo a first order
phase transition into the normal state. Considering the idea that it is the
escape of high energy phonons that governs the recovery of the
superconducting state \cite{RothwarfTaylor}, and following the experimental
work of Sai-Halasz at al., who failed to observe the predicted first-order
phase transition \cite{SaiHalasz}, Parker suggested an alternative model 
\cite{Parker}. He argued that in out-of-equilibrium the QPs and the
condensate\ share the same chemical potential. However, their temperature T$%
^{\ast}$, which is the same as the temperature of $\hbar \omega>2\Delta$
phonons, differs from the temperature of low energy ($\hbar\omega<2\Delta$)
phonons (base temperature). While it was argued that the above mentioned $%
T^{\ast}$ and\ $\mu^{\ast}$ models may not adequately describe the behavior
in the vicinity of the phase transition \cite{Elesin}, the solutions of
kinetic equations clearly show non-equilibrium effects in phonon population,
consistent with the idea put forward by Parker (see e.g. Figure 8 of Ref. 
\cite{Elesin}).

\subsection{Enhancement of superconductivity with sub-gap excitation}

\label{EnhancHistory}

As mentioned above, studies of superconducting properties under continuous
irradiation with sub-gap radiation (e.g. microwaves) showed surprising
results \cite{Wyatt,Dayem,Klapwijk,Clarke}. These are illustrated in Figure
2, summarizing observations of microwave enhancement of superconductivity in
superconducting thin-film strips of Al \cite{Klapwijk77}. It was
demonstrated that at temperatures close to the equilibrium critical
temperature and for frequencies of microwave radiation above some
characteristic value, one can observe a measurable increase in the
superconducting critical current (Fig. 2a), as well as an increase in the
superconducting critical temperature and the magnitude of the
superconducting gap. In these experiments, the order parameter jumped
discontinuously to zero at the superconducting transition (see Fig. 2b) \cite%
{Klapwijk77,Schmid77}. With microwave irradiation, superconductivity could
be observed above the equilibrium critical temperature, a phenomenon that is
commonly referred to as \emph{photo-induced superconductivity}.

\begin{figure}[ptb]
\includegraphics[width=120mm]{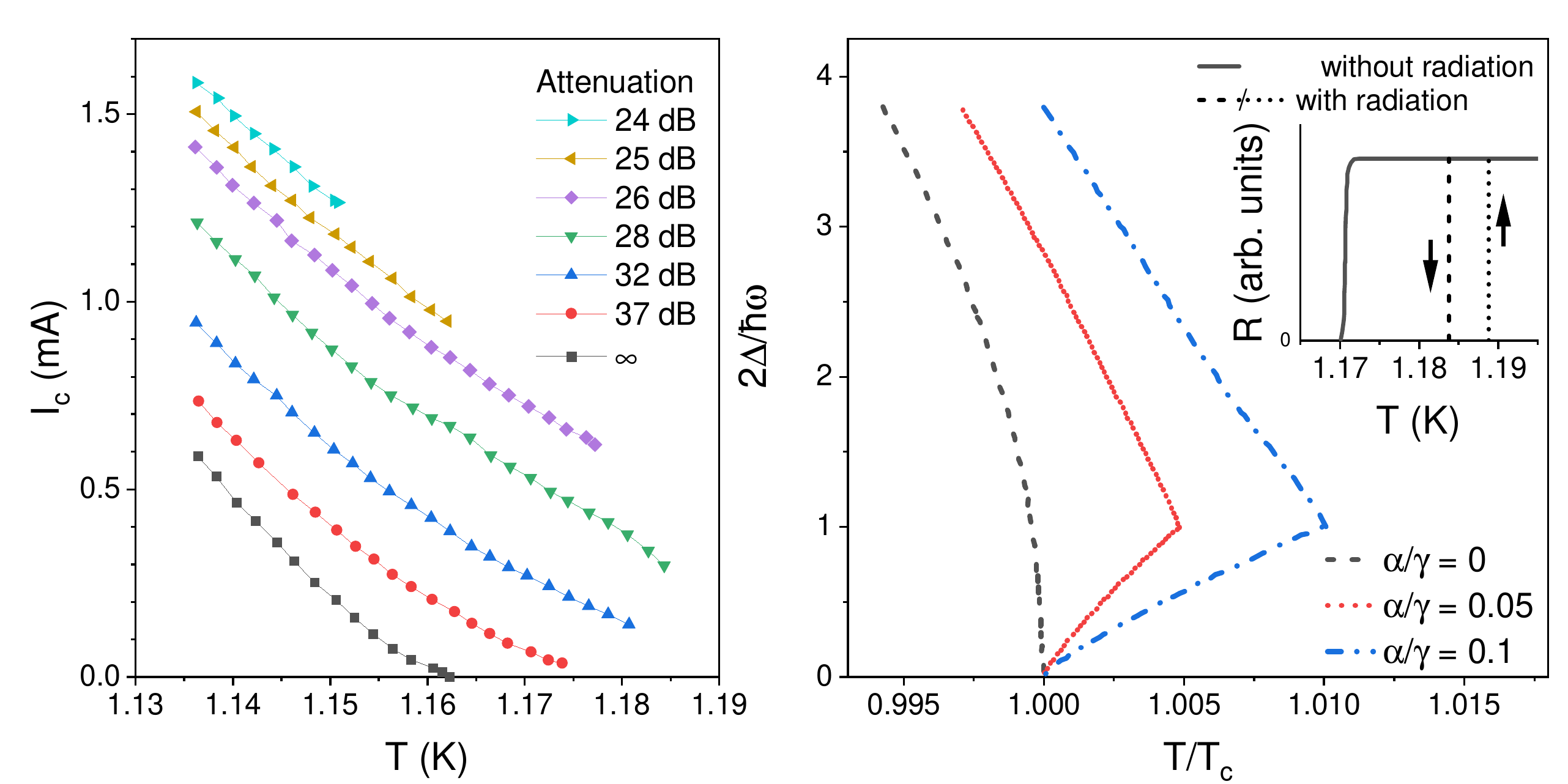} 
\caption{Temperature dependence of the critical current for various levels
of microwave power at 3 GHz (attenuations are shown on the plot, with $%
\infty $ corresponding to the absence of microwave radiation). The results
show not only an enhanced critical current, but also an increase in the
superconducting transition temperature. In equilibrium T$_{c}=1.170$ K,
while under microwave radiation the sample superconducts already at 1.184 K.
(b) The calculated temperature dependence of the superconducting gap in Al
under 3 GHz radiation using the Eliashberg theory. Inset presents the
resistivity curves with and without microwave radiation (for the later, the
hysteresis implies the first order phase transition under microwave
excitation). The presented data were adapted from Ref. \protect\cite%
{Klapwijk77}.}
\end{figure}

These observations were explained by Eliashberg \cite%
{Eliashberg70,Eliashberg73}, who realized that in the vicinity of the
critical temperature, the effect is mainly related to a non-equilibrium
distribution of quasiparticles induced by the microwave field. In
particular, at finite temperatures, sub-gap radiation couples to thermally
excited QPs at the gap edge, giving rise to excitation of quasiparticles to
energies further from the gap edge. Within the BCS theory, the
superconducting gap is determined by the self-consistent
temperature-dependent gap equation:

\begin{equation}
\frac{1}{N(0)V_{0}}=\int_{0}^{\hbar\omega_{c}}d\xi\frac{\tanh\big{(}\frac {1%
}{2}\beta\sqrt{\xi^{2}+|\Delta|^{2}}\big{)}}{\sqrt{\xi^{2}+|\Delta|^{2}}},
\label{bcs_gap}
\end{equation}
where $\xi$ is the kinetic energy relative to the Fermi level. The maximum
gap enhancement would result by exciting thermally excited QPs as far as
possible from the gap-edge. Using Eq.(\ref{bcs_gap}) and expanding for small
changes of the order parameter $|\Delta|^{2}=|\Delta_{0}|^{2}(1+\delta%
\Delta/|\Delta _{0}|)$ as a result of changes in the Fermi-Dirac
distribution, with $f(\xi)=f_{FD}(\sqrt{\xi^{2}+|\Delta|^{2}})+\delta f(\xi)$%
, where $f_{FD}$ is the Fermi-Dirac distribution and $\delta f(\xi)$ is the
excitation induced change in the distribution function, one obtains (in the
limit of small relative gap changes) \cite{Rousseau} 
\begin{equation}
\frac{\delta\Delta}{|\Delta_{0}|}\approx-C\int_{0}^{\hbar\omega_{c}}d\xi 
\frac{\delta f(\xi)}{\sqrt{\xi^{2}+|\Delta_{0}|^{2}}}.
\label{eq:gap_increase}
\end{equation}
Here $C$ is a positive dimensionless constant that generally depends on the
base temperature. From Eq.(\ref{eq:gap_increase}) it follows that pushing
the gap-edge quasiparticles to higher energies, $\delta f(0)<0$, results in $%
\delta\Delta>0$.

Since these experiments were performed using continuous microwave radiation,
while inducing changes in the distribution functions require slow
thermalization of QPs, it is no surprise that enhancement was observed only
for pump frequencies exceeding the characteristic QP scattering rate.
Moreover, the observed enhancements of the gap and critical temperature were
quite small (e.g. an increase of $T_{c}$ from $1.170$ K to $1.184$ K in Al -
see Figure 2).

Detailed experimental studies as a function of frequency, power and base
temperature showed some departures from the Eliashberg theoretical
predictions \cite{Schmid77,Schoen}. Chang and Scalapino \cite%
{Scalapino,Scalapino2,Chang78} proposed an extension of the\ Eliashberg
theory by considering the energy dependence of the QP recombination rate (in
Fig. 1a the recombination rate R is considered to be a constant, but in
general it depends on QP energy). The high energy quasiparticles recombine
into Cooper pairs faster, since there are more available phonon states to
relax the energy to. By analyzing the coupled Boltzmann equations for phonon
and quasiparticle distributions under sub-gap excitation \cite{Chang78},
Chang and Scalapino demonstrated cooling of a superconductor upon
photo-excitation.

Finally, not only microwave radiation, but also excitation with sub-gap
microwave phonons was demonstrated to give rise to enhancement of
superconductivity \cite{Tredwell75,Tredwell76,Seligson}, which could be
explained within the generalized Eliashberg model \cite{Cirillo}.

For further reading on experimental and theoretical approaches to realize
enhancement of superconductivity using quasi-equilibrium sub-gap excitation
we refer to several reviews to \cite{Pals,Mooij}.

\section{Real-time spectroscopy of the superconducting gap dynamics \label%
{realTimeSpectroscopy}}

To study the dynamics of light-induced suppression and recovery of the
superconducting state in real-time the common approach is to use the
so-called pump-probe spectroscopy \cite{Hilton}. In the simplest version,
femtosecond optical pulses of the same photon energy are used for excitation
(pump) and for probing the induced changes in optical constants (probe) as a
function of time delay between the two. Most of the early studies were
performed in this configuration \cite%
{Han,Chwalek,Albrecht,Stevens,DemsarYBCO,kabanov,Segre,mercury,Gedik,Kusar,Giannetti09,Mertelj,Chia,Mansart,GedikPnic}%
. While the photo-induced changes at optical (probe) frequencies do reflect,
via Kramers-Kronig relations, the photo-induced changes and dynamics of the
low-energy excitation spectrum (e.g. the superconducting gap), probing the
gap dynamics using far-infrared (THz) pulses \cite%
{Federici,Kaindl,THzAveritt,Carr,MgB2,Lobo,Beck1,KaindlBiSCO,Pashkin,Beyer,PCCOBeck,Sindler}
or with photoemission \cite%
{Perfetti,Uwe,Lanzara,ZXShen,HeHRArpes,Dessau,Bauer}, is much more
straightforward to interpret.

In the following, we first focus on phenomena observed on a conventional BCS
superconductor NbN obtained using time-resolved THz spectroscopy \cite{Beck1}%
. We note that similar results were obtained in another conventional BCS
superconductor MgB$_{2}$ \cite{MgB2}, as well as in an electron-doped
cuprate superconductor Pr$_{2-x}$Ce$_{x}$CuO$_{4}$ \cite{PCCOBeck}. The
latter is especially interesting, given the d-wave symmetry of the order
parameter in cuprates, as elaborated in section \ref{cuprates}.

Before discussing the real-time dynamics of superconductors driven by pulsed
laser excitation, we should briefly address the characteristic excitation
densities. To do so, let us assume a conventional BCS s-wave superconductor,
where most of the absorbed energy of optical pulse is used to break Cooper
pairs (as elaborately discussed in the following sections this condition
seems to be fulfilled in conventional superconductors, while in the high-T$%
_{c}$ cuprate superconductors the situation may be more complicated). Under
this assumption, the absorbed energy density, \emph{A}, should be lower than
the superconducting condensation energy, $E_{c}$, if we are to discuss the
dynamics in the perturbative regime (for $\emph{A}\approx E_{c}$ excitation
results in melting of superconductivity). Here \emph{A }= $\frac {F(1-R)}{%
l_{opt}}$, where F is the fluence of the incoming optical pulse, typically
in the $\mu$J/cm$^{2}$ range, R is the reflectivity of the sample and $%
l_{opt}$ is the optical penetration depth. The condensation energy, $E_{c}$
, one the other hand, is - for a conventional BCS s-wave superconductor -
given by $E_{c}$ $=B_{c}^{2}/2\mu_{0}=N(0)\Delta^{2}/2$, where $B_{c}$ is
the thermodynamical critical field and N(0) is the normal state density of
states. The above condition, however, does not yet imply the dynamics to be
linear, i.e. excitation density independent. Since the relaxation processes
contain bi-molecular terms (see Eq. \ref{bimolecular}), the recovery
dynamics will be linear only for excitation densities, where the density of
photoexcited quasiparticles is small compared to the density of thermally
excited quasiparticles, given by $n_{T}\simeq N(0)\sqrt{2\pi\Delta k_{B}T}%
\exp(-\Delta/T)$.

\subsection{Photo-induced dynamics of the superconducting gap in NbN}

Thin NbN thin films on MgO substrates, with T$_{c}$ of 15.4 K were
investigated using near-infrared pump THz-probe spectroscopy \cite{Emitter}.
Following linear THz spectroscopy studies (Figure 3a), where the temperature
dependence of the superconducting gap, $\Delta(T)$, is extracted using the
fit to the complex optical conductivity \cite{Zimmermann}, $\sigma(\omega)$,
the dynamics of the gap following excitation with a 50 fs near-infrared (800
nm) pulse can be studied. The time evolution of the SC state as a function
of time delay, $t$, after optical excitation, was studied by either directly
measuring spectrally resolved $\sigma(\omega,t)$ or by measuring the
spectrally integrated response by tracking the time evolution of the induced
changes in the transmitted electric field, $\Delta E_{tr}(t^{\prime},t)$, at
a fixed $t^{\prime}$ (see Fig. 3c). The latter approach is particularly
useful for studying dynamics at low excitations. Given the fact that films
are optically thin, they are homogeneously excited. The absorbed energy
density, \emph{A }(in units meV per unit cell volume, meV/ucv), can be
determined from the reflectivity and transmission at 800 nm (pump).

\begin{figure}[ptb]
\includegraphics[width=120mm]{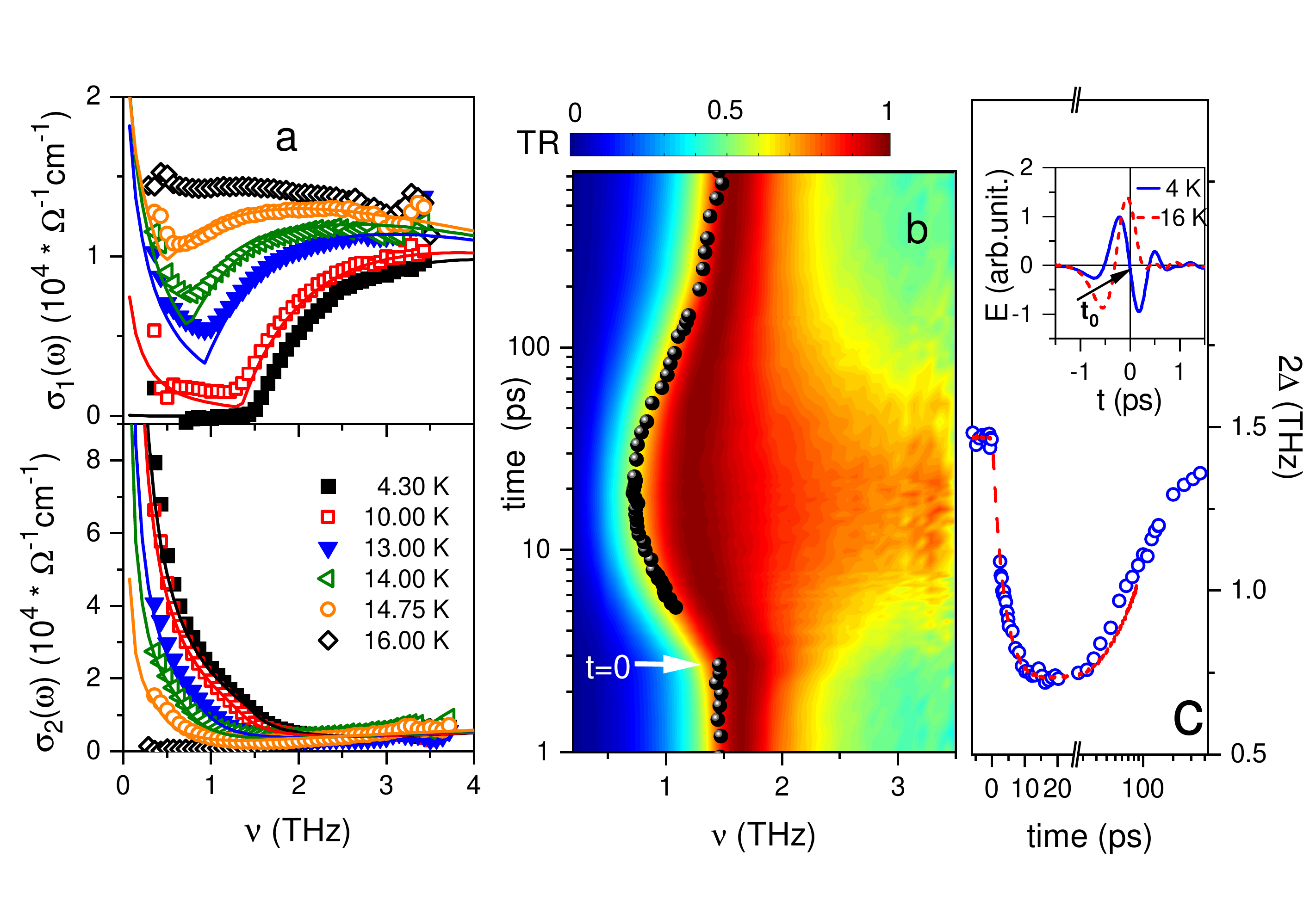} 
\caption{(a) The T-dependence of complex optical conductivity, $\protect%
\sigma\left( \protect\omega\right) $, of 15 nm NbN film (T$_{c}=15.4$ K).
Solid lines are fits with the BCS equations \protect\cite{Zimmermann},
providing access to the T-dependence of the gap, $\Delta$(T). (b) The time
(t) evolution of the gap (black spheres), determined at each time delay,$t$
, from $\protect\sigma\left( \protect\omega,t\right) $. The background
contour-plot presents the corresponding ratio of the THz transmission in the
superconducting and normal state, normalized to its peak value. It reflects
the time-dependence of the gap. (c) Presents the gap dynamics, obtained in
two configurations: open circles present the values obtained from the
analysis of spectrally resolved $\protect\sigma(\protect\omega,t)$, while
the dashed line presents the data obtained using spectrally integrated
approach. The later, given by the dashed line, is obtained by monitoring the
induced changes in the electric field transient at a fixed point of the
transmitted THz electric field transient ($t^{\prime}=0$ in the inset).}
\end{figure}

There are several interesting observations in Fig. 3 that should be
addressed:

(a) The time-evolution of $\sigma(\omega)$ shows that even at quite high
excitation densities as presented here, $\sigma(\omega,t)$ can be well
described by the time-evolution of the superconducting gap, $\Delta(t)$. In
fact, the data suggest that $\sigma(\omega,t)$ can be approximated by $%
\sigma(\omega,T^{\ast}(t))$, where at any given time delay (apart from
perhaps on the sub-ps timescale) the superconducting state can be
approximated by an effective temperature $T^{\ast}$, similar to the early
description by Parker \cite{Parker}.

(b) Despite the fact that the sample is excited by a 50 fs optical pulse,
the pair-breaking process (initial decrease of $\Delta$) proceeds on a long
10 ps timescale. The observation is surprising, since the $e-e$
thermalization time in metals is assumed to proceed on a sub-ps timescale 
\cite{AllenEx,BrorsonEx,BookChapter}. Thus, one would naively expect the
Cooper-pair breaking to be finished within a few 100 fs. The observation of
slow pair-breaking suggests an alternative scenario, where bosonic
excitations (e.g. phonons), generated during the initial decay of hot
carriers towards the gap-edge, are responsible for pair-breaking.

(c) Finally, in NbN the superconducting state is almost completely recovered
on a timescale of a few hundred ps. While the timescale is much shorter than
the SC recovery timescales deduced from continuous excitation experiments on
Al or Sn \cite{MillerDayem,Gray69}, this observation does not prove the
early estimates wrong. Based on available data from time-resolved studies on
different superconductors, ranging from conventional BCS superconductors to
high-T$_{c}$ cuprates, one can conclude that the characteristic SC recovery
timescale is roughly inversely proportional to the size of the
superconducting gap.

The most interesting and insightful is however following changes in dynamics
as a function of base temperature and excitation density. Here, the data
reveal that both pair-breaking rate and the superconducting state recovery
timescale depend on both, the excitation density and temperature \cite%
{MgB2,Beck1,PCCOBeck,Sindler}. These dependencies provide access to the
nature of the underlying processes and material specific microscopic
parameters illustrated in Fig. 1.\newline

From Figure 3 it follows that in NbN \cite{Beck1,Sindler} (the same is true
for MgB$_{2}$ \cite{MgB2} and Pr$_{2-x}$Ce$_{x}$CuO$_{4}$ \cite{PCCOBeck}),
the Cooper pair-breaking process ($\approx10$ ps) and SC state recovery ($%
\approx100$ ps) are well separated in time, which allows for a detailed view
of the underlying microscopic processes based on their temperature and
excitation density dependencies. Below we address the two processes in more
detail.

\subsubsection{Cooper pair breaking dynamics: Determination of the
electron-phonon coupling}

\label{NbNBreaking}

The rise-time dynamics is clearly reflecting the pair-breaking processes
(i.e. the initial reduction of condensate density and an increase in the
density of QPs). The pair-breaking was found to be temperature dependent
(the process is faster as T is increased) \cite{MgB2}, and photo-excitation
density dependent (see Figure 4) \cite{MgB2,Beck1}. Especially the latter
provides strong indications that the delayed pair-breaking is caused by
high-frequency phonons, released during the sub-ps relaxation of high energy
electrons \cite{BookChapter,KabanovTTM,Obergfell}.

\begin{figure}[ptb]
\includegraphics[width=120mm]{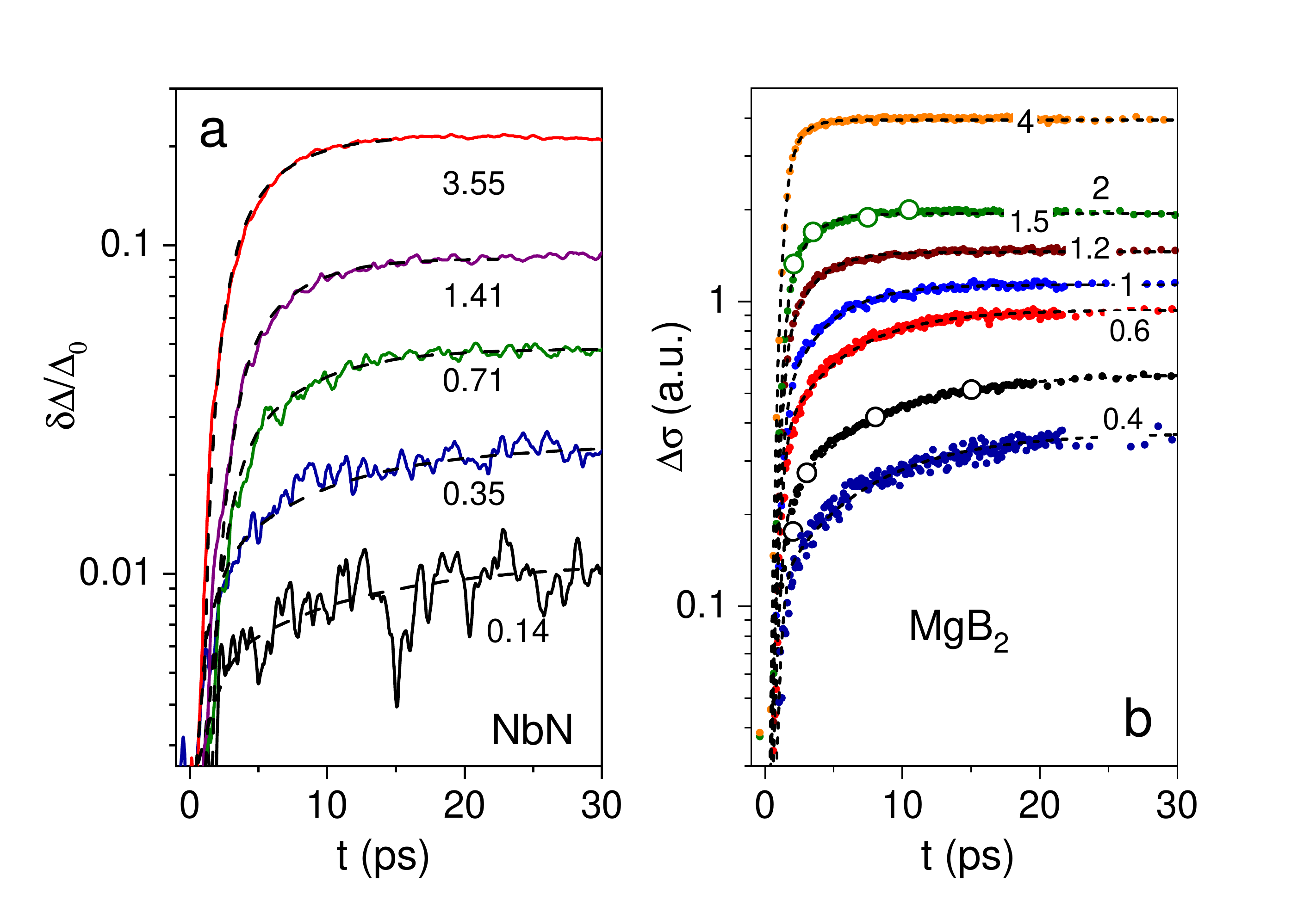} 
\caption{(a) Superconducting depletion dynamics shown as the relative change
of the gap, $\protect\delta\Delta/\Delta_{0}$, recorded at 4.3 K on NbN film
for various excitation densities in mJ/cm$^{3}$ \protect\cite{Beck1}. The
dashed lines are fits to the data with Eq.(7). (b) Superconducting depletion
dynamics 7 K taken at various fluences (in $\protect\mu$J/cm$^{2}$) on MgB$%
_{2}$ \protect\cite{MgB2}. Solid circles represent the data obtained by
energy integrated approach (see Fig. 3b), while the open circles correspond
to $\Delta\protect\sigma(\protect\omega)$ measured directly. Again, the
dashed lines are fits to the data using Eq.(7).}
\end{figure}

As shown in Figure 4, in both NbN and MgB$_{2}$, at low excitation densities
the pair-breaking can take tens of ps, speeding up upon increasing the
excitation density. To account for these observations, in particular the
excitation density dependence of the pair-breaking process, we consider the
interplay between the condensate/QPs and high-frequency ($%
\hbar\omega>2\Delta $) phonons as proposed by Rothwarf and Taylor \cite%
{RothwarfTaylor}. Within the model, the dynamics of the QP and high
frequency phonon (HFP) densities, $n$ and $N$, are described by a set of two
coupled differential equations \cite{RothwarfTaylor}: 
\begin{equation}
dn/dt=I_{0}+\eta N-Rn^{2}\text{ \ and }dN/dt=J_{0}-\eta
N/2+Rn^{2}/2-\gamma(N-N_{T})  \label{RTeq}
\end{equation}

Here $\eta$ is the probability for pair-breaking by HFP absorption, $R$ is
the bare QP recombination rate with the creation of a HFP, $N_{T}$ is the
concentration of HFP in thermal equilibrium at temperature $T$, and $\gamma$
their decay rate. The rate $\gamma$ could be governed\ either by anharmonic
decay of HFP \cite{kabanov}, since $\omega<2\Delta$ phonons do not have
sufficient energy to break Cooper-pairs, or by the diffusion of HFP into the
substrate \cite{escape}. Finally, $I_{0}$ and $J_{0}$ represent the initial
densities of excess QPs and HFPs, which are generated during the avalanche
relaxation of hot QPs towards the gap edge.

Now, let us imagine that the decay of HFP is the limiting step, governing
the recovery of SC, i.e. the so-called phonon-bottleneck scenario by
Rothwarf and Taylor \cite{RothwarfTaylor}. Since the SC recovery dynamics
proceed on a much longer timescale (100's of ps), one can - in the first
approximation - neglect the decay of HFP given by the last term in Eq.(\ref%
{RTeq}), to reduce the equations to 
\begin{equation}
dn/dt=I_{0}+\beta N-Rn^{2}\text{ \ and }dN/dt=J_{0}+\frac{1}{2}[Rn^{2}-\beta
N].  \label{RT}
\end{equation}
In this case, the early stage of relaxation will result in thermalization of
QPs and HFP, which will reach a quasi-equilibrium state, described by a
detailed balance equation, $Rn_{T^{\ast}}^{2}=\beta N_{T^{\ast}}$, where $%
T^{\ast}$ will be some new effective temperature (as in the model proposed
by Parker \cite{Parker}). For a SC gap much smaller than the Debye energy
and the phonon density of states $D(\omega)=9\nu\omega^{2}/\omega_{D}^{3}$
the HFP density is given by $N_{T}=\frac{36\nu\Delta^{2}T}{\omega_{D}^{3}}%
\exp(-2\Delta/k_{B}T)$ \cite{kabanov}. On the other hand, the QP density is
given by $n_{T}\simeq N(0)\sqrt{2\pi\Delta k_{B}T}\exp(-\Delta/T)$. This
means that at temperatures, where $k_{B}T\ll\Delta$, in equilibrium most of
the energy is actually in the QP channel (note the $2\Delta/k_{B}T$ vs $%
\Delta/T$ arguments in the exponents). Moreover, from the detailed balance
equation it follows that $\eta/R=$ $n_{T}^{2}/N_{T}=\frac{%
N(0)^{2}\pi\omega_{D}^{3}}{18\nu\Delta}$.

It is instructive to investigate the solutions of Eqs.(\ref{RT}) for
different excitation conditions. For simplicity, we can consider the low
temperature limit, where the density of thermally excited QP's and HFP can
be neglected. Secondly, we note that the ratio of microscopic constants $%
\eta/R$ has the dimensionality of concentration. Hence, one can introduce
dimensionless QP and HFP concentrations, $q\equiv Rn/\eta$ and $p\equiv
RN/\eta$, while $\theta\equiv\eta t$ is the dimensionless time \cite{RT}.
With this, Eqs.(\ref{RT}) read: 
\begin{equation}
dq/d\theta=p-q^{2}\text{ \ and }dp/d\theta=-p/2+q^{2}/2\text{,}
\label{RTdim}
\end{equation}
with the initial conditions $p(0)=p_{0}$, $q(0)=q_{0}$. These coupled
equations have analytic solutions $q(\theta)$ and $p(\theta)$ \cite{MgB2,RT}%
. Here we limit the discussion to the dynamics of QP density, $q(\theta)$,
which can be directly compared to the experimental data $\Delta(t)$ in
Figure 4. The analytic solution for $q(\theta)$ is

\begin{equation}
q(\theta)=\left[ -\frac{1}{4}-\frac{\xi^{-1}}{2}+\frac{\xi^{-1}}{1-K\exp{%
(-\theta/\xi)}}\right] ,  \label{Q_RT}
\end{equation}
where $\xi^{-1}=\sqrt{1/4+4p_{0}+2q_{0}}$ and $K=\frac{(4q_{0}+1)-2\xi^{-1}}{%
4q_{0}+1)+2\xi^{-1}}$ are dimensionless constants, depending on the initial
conditions ($p_{0},$ $q_{0}$). For long time delays, $\theta\rightarrow%
\infty $, it follows that $q^{2}(\infty)=p(\infty)$ and $q(\infty)=\frac{1}{4%
}(\sqrt{1+16p_{0}+8q_{0}}-1)$ .

It is further instructive to take a closer look at the meaning of the
dimensionless parameter $K$. Obviously, $K=0$ corresponds to the case when
immediately after photo-excitation the stationary solution is realized, i.e.
when $q_{0}^{2}=p_{0}=q^{2}(\infty)=p(\infty)$. In this case the time
evolution of the QP density, or of the reduction of $\Delta$, would be a
step function. The regime $0<K\leq1$ corresponds to the situation when $%
q_{0}>q(\infty)$, while $-1\leq K<0$ represents the $q_{0}<q(\infty)$. For
the latter case, $p_{0}>p(\infty)$, which results in an additional
pair-breaking until $q^{2}(\infty)=p(\infty)$ is reached.

Figure 5 presents a few simulations of $q(\theta)$ which illustrate
different limiting cases. Note that the function (\ref{Q_RT}) was convoluted
with the Heaviside step-function with the rise-time of 100 fs, which
accounts for photo-excitation and avalanche initial
relaxation/multiplication process of hot quasiparticles. We consider a
superconductor with $\Delta=4$ meV and the absorbed energy density \emph{A} [%
$\mu eV/$ucv] distributed between QPs and HFPs such that $p_{0}=\frac{R}{\eta%
}\frac{\emph{A}(1-x)}{2\Delta}$ and $q_{0}=\frac{R}{\eta}\frac{\emph{A}x}{%
\Delta}$, where $x$ is the fraction of the absorbed energy density in the QP
channel ($1\geq x\geq0$).

\begin{figure}[ptb]
\includegraphics[width=120mm]{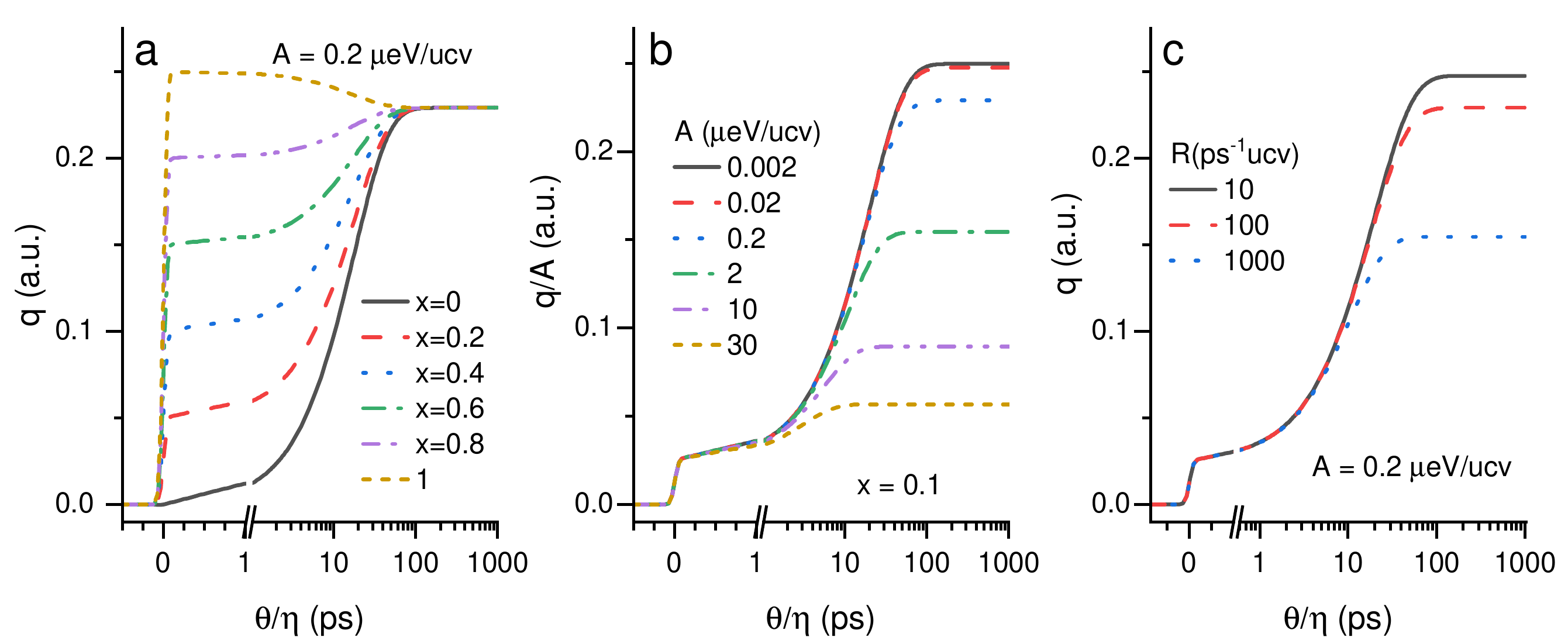} 
\caption{Simulation of the pair-breaking dynamics using Eq.(\protect\ref%
{Q_RT}) in the $T\rightarrow0$ K limit (also valid when photo-excited
carrier density much higher then the thermal one). For all simulations $%
\Delta=4$ meV and the pair breaking rate is $\protect\eta=0.1$ ps$^{-1}$.
Panel (a) presents the dynamics for constant excitation density, R = 100 ps$%
^{-1}$ucv and different fractions of the absorbed energy in the QP channel ($%
0<x<1$). (b) Excitation dependence of dynamics for $x=0.1$ and R = 100 ps$%
^{-1}$ucv, note that the signal is normalized to excitation density. (c) The
dependence of dynamics on the recombination rate and at constant \emph{A}. }
\end{figure}

Figure 5a displays the evolution of the QP density for a constant excitation
density and varying $x$, demonstrating the slowly rising signal due to the
excess density of HFP (compared to $p(\infty)$). It is noteworthy that only
for $x\rightarrow1$ the situation is realized, that the QP density is
decaying with time in this so-called \textit{pre-bottleneck regime}. This
actually reflects the ratio of $n_{T}\propto e^{-\Delta/k_{B}T}$ and $%
N_{T}\propto e^{-2\Delta/k_{B}T}$ discussed above. In other words, for $%
k_{B}T<<\Delta$ in quasi-equilibrium $n_{T}\gg N_{T}$, and nearly all of the
absorbed energy ends up in the QP channel. Therefore it is no surprise, that
even if only 20\% of the absorbed energy is initially in the HFP channel,
the delayed buildup of QP density (or suppression of $\Delta$) can be
observed.

Panel (b) presents the dependence of pair-breaking on excitation density,
with all other parameters being constant. With signal normalized to \emph{A}
the dependence of the apparent pair-breaking time on \emph{A} reveals the
behavior observed in experiments (see Fig. 4). Sub-linearity in the value of 
$q(\infty)$, i.e. the photo-induced reduction of the gap, at high excitation
densities is also revealed. Finally, panel (c) presents the evolution of the
dynamics when varying the recombination rate (R) while keeping the
pair-breaking rate $\eta$ constant.

Using this analysis on NbN (the fits to the data are shown by dashed lines
in Figure 4a, where $\Delta$ is known, and it is demonstrated (see section %
\ref{NbNEnerg}) that the absorbed energy can indeed be simply distributed
between the QPs and HFP, one can extract $\eta$ and $K$ and the fraction of
the energy density initially stored in the QP channel $x$. The best fit is
obtained when $x=0.09$, giving the values of the microscopic constants $%
\eta^{-1}=6\pm1$ ps, $R=$ $160\pm20$ ps$^{-1}$unit cell.

Since the pair-recombination rate $R$ can be expressed in terms of
microscopic parameters through $R=\frac{8\pi\lambda\Delta^{2}}{\hbar N\left(
0\right) \Omega_{c}^{2}}$ \cite{mercury,Ovchinikov}, where $\Omega_{c}$ is
the phonon cut-off frequency and $N(0)$ is the density of states, the value
of $\lambda$ can be determined. Taking the known values for $\Delta$, $N(0)$
and\ $\Omega _{c}=16$ THz \cite{Geibel}, we obtain $\lambda=1.1\pm0.12$,
which is in excellent agreement with the theoretical estimates, $%
\lambda=1-1.12$ \cite{Weber,lambda}. Importantly, such an approach does not
suffer from the questionable assumptions \cite%
{BookChapter,KabanovTTM,Obergfell} of the so-called two-temperature model,
the model commonly used to determine $\lambda$ from time-resolved
experiments \cite{BrorsonEx,allen}.

\subsubsection{Energetics of the gap suppression in conventional
superconductors}

\label{NbNEnerg}

As addressed above, when at $k_{B}T<<\Delta$ the absorbed energy is
distributed between the QPs and HFP and the quasi-equilibrium is reached,
most of the energy is in the QP system. This fact provides the means to
estimate the condensation energy of a superconductor.

\begin{figure}[ptb]
\includegraphics[width=120mm]{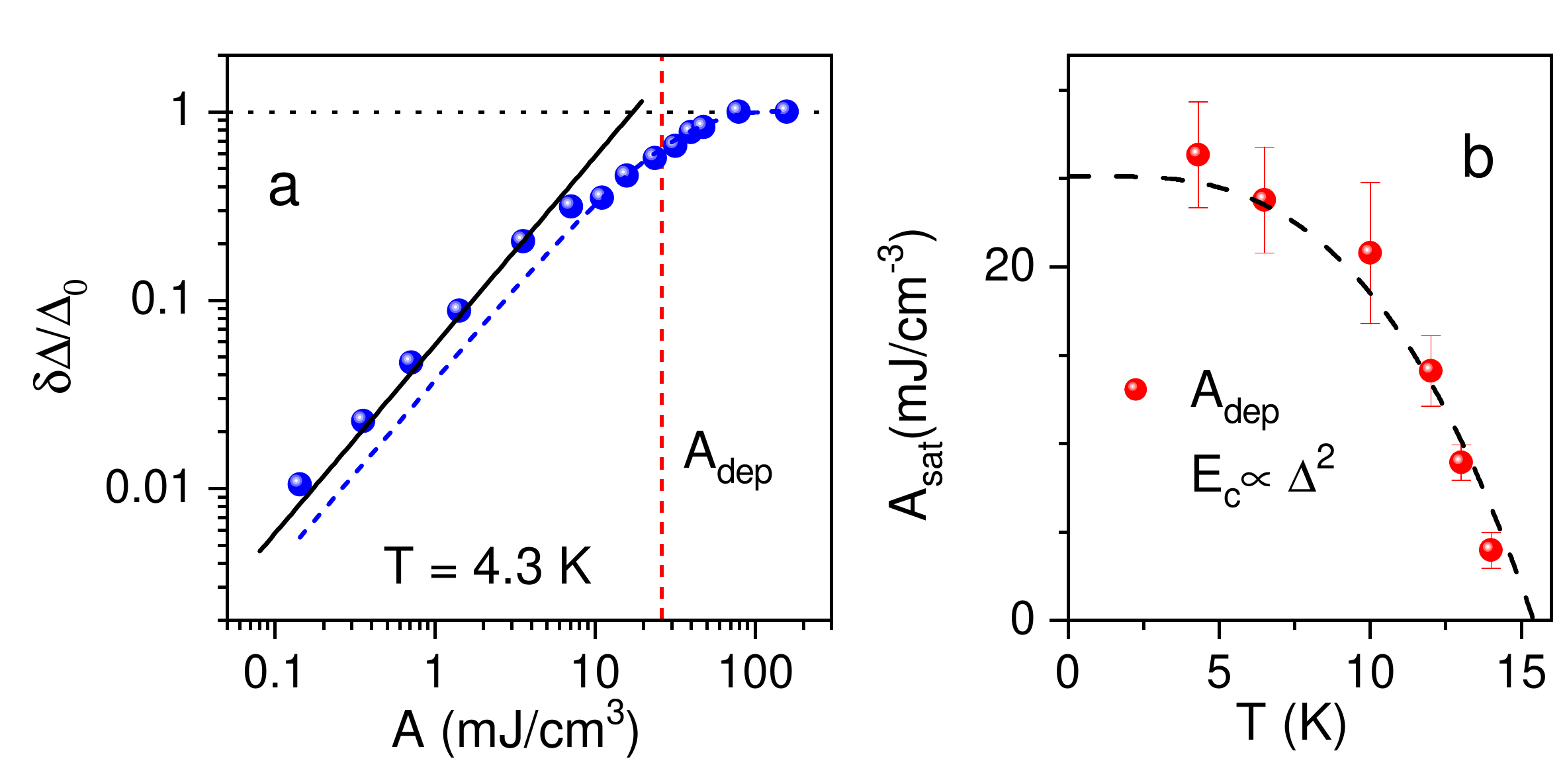} 
\caption{ (a) The dependence of the peak gap suppression, $\protect\delta%
\Delta /\Delta_{0}$, in NbN on \emph{A } at 4.3 K (see the traces in Figure
4a). The dashed line is a fit to the simple saturation model, the solid line
is the linear fit. (b) The T-dependence of the energy density required to
deplete superconductivity, $A_{dep}$, compared to the T-dependence of $%
\Delta^{2}$ (dashed black line).}
\end{figure}

Figure 6a presents the dependence of the (maximal) relative change of the SC
gap ($\delta\Delta/\Delta$) in NbN recorded at 4.3 K as a function of
absorbed energy density (in mJ/cm$^{3}$). At low excitation densities $%
\delta \Delta/\Delta$ increases linearly with \emph{A}, followed by a
saturation (plateau at $\delta\Delta/\Delta=1$ is reached when
superconductivity is completely suppressed). Following the discussion of the
coupled QP and HFP system above, it becomes clear that for the complete
collapse of the gap, where QPs would be in a quasi-thermal equilibrium with
HFP, the required energy would equal the thermal energy to heat up the
sample from the base temperature to above T$_{c}$. This follows from the
fact that -- in the absence of the gap - all phonons are the HFP (the energy
stored in the HFP system in this case will be dominant). On the other hand,
the rate at which the gap (or condensate density) gets suppressed in the
limit of weak perturbations, should provide an estimate of the condensation
energy. Alternatively, this may be particularly relevant for bulk samples
where the absorbed energy density varies with the depth, an estimate of the
characteristic $A$ required to suppress SC, one commonly uses a
phenomenological saturation model, $\delta\Delta/\Delta_{0}=$ $\left(
1-\exp\left( -A/A_{dep}\right) \right) $. Here $A_{dep}$ is the
characteristic absorbed energy density required to deplete SC. It follows
from Figure 6a, that for NbN $A_{dep}\left( 4.3\text{ K}\right) \approx25$
mJ/cm$^{3}$. Within the experimental accuracy, $A_{dep}\left( 4.3\text{ K}%
\right) \approx25$ mJ/cm$^{3}$ is comparable to the thermodynamic SC
condensation energy, $E_{c}$, given by $E_{c}=B_{c}^{2}/2\mu_{0}=N(0)%
\Delta^{2}/2$. Here $B_{c}$ is the thermodynamic critical field and $N(0)$
is the electronic density of states at the Fermi level. For NbN $B_{c}=0.234$
T \cite{Geibel}, giving $E_{c}=22$ mJ/cm$^{3}$, nearly identical to $%
A_{dep}\left( 4.3\text{ K}\right) $. Moreover, as shown in Figure 6b, the
T-dependence of $A_{dep}$ is found to follow the T-dependence of $\Delta^{2}$
(dashed line). We note that the observation in NbN is in strong contrast to
similar studies in cuprates \cite{Kusar,Pashkin,Beyer,PCCOBeck}, where $%
A_{dep}$ is found to be about one order of magnitude higher than $E_{c}$ and
is nearly independent on temperature \cite{Beyer,PCCOBeck}.

\subsubsection{Superconducting state recovery dynamics at temperatures far
below T$_{c}$}

\label{NbNRecovery}

Now, let us turn our attention to the dependence of the SC recovery time on
temperature and excitation density, which is for NbN shown in Figure 7.

\begin{figure}[ptb]
\includegraphics[width=120mm]{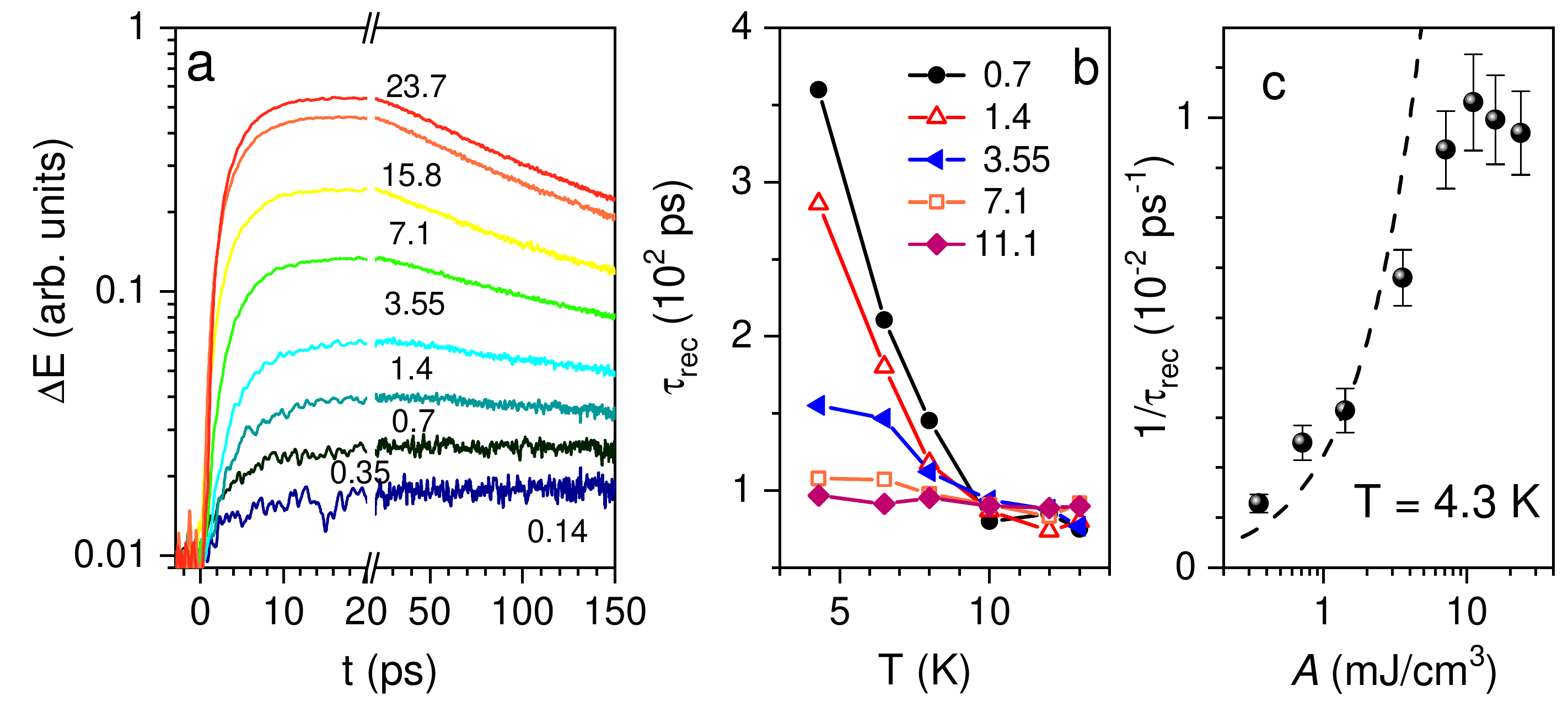} 
\caption{(a) The time evolution of the photoinduced gap suppression
(spectrally integrated response - see text) recorded on NbN thin film at 4.3
K with excitation density covering over two orders of magnitude. (b)
Temperature and excitation dependence of the superconducting state recovery
time, $\protect\tau_{rec}$, obtained by a fit of the data in panel (a) with
an exponential decay function. (c) The low temperature superconducting state
recovery rate in NbN as a function of excitation density.}
\end{figure}

Fig. 7a presents the relaxation dynamics recorded at 4K over 2 orders of
magnitude in excitation density. While at the lowest excitation densities
the recovery time, $\tau_{rec},$ substantially exceeds the data acquisition
window, one can clearly see speeding up of relaxation dynamic upon
increasing excitation density. As shown in Fig. 7b the dynamics is below $%
\approx10$ K both temperature and excitation density dependent.

As shown in Fig. 7c, at 4 K $\tau_{rec}^{-1}$ first increases linearly with 
\emph{A}, mimicking the intrinsic bi-molecular kinetics of the QP
recombination discussed in section \ref{DynHistory}. However, given the fact
that the early time dynamics in NbN reveals a delayed pair-breaking with
build-up of quasi-equilibrium between QPs and HFP (see section \ref%
{NbNBreaking}), it is clear that the recovery dynamics is determined by the
decay of\ the HFP population. I.e., the SC recovery dynamics should be
described within the phonon bottleneck scenario of Rothwarf and Taylor \cite%
{RothwarfTaylor,RT}.

Before addressing this, we should note that such excitation density and
temperature dependence of SC state recovery was first observed in cuprate
superconductors \cite{KaindlBiSCO,GedikYBCO,Hinton}, where it was argued to
be an evidence for the absence of phonon (boson) bottleneck. It was stated
that in the case of a phonon bottleneck the relaxation dynamics should be,
at low temperatures, roughly constant and excitation density independent 
\cite{GedikYBCO,Orenstein}. Surprisingly, and perhaps counter-intuitive,
considering full solution of the coupled Rothwarf-Taylor equations, Eqs.(\ref%
{RTeq}), it has been demonstrated, that even in the case of
phonon-bottleneck the relaxation should display excitation density and
temperature dependence \cite{RT} as observed in NbN (Figure 7).

While for the complete derivation of analytic solutions for SC state
recovery we refer to Ref. \cite{RT}, we briefly sketch the main arguments
here. We remain with the description using dimensionless quantities as
above. First of all, including the phonon decay term, the Rothwarf-Taylor
equations read 
\begin{equation}
dq/d\theta=p-q^{2}\text{ \ and }dp/d\theta=-p/2+q^{2}/2-\tilde{\gamma}%
(p-p_{T})\text{,}  \label{RTfullDim}
\end{equation}
where $\tilde{\gamma}\equiv\gamma/\eta$ is the dimensionless HFP decay rate
and $p_{T}$ is the thermal density of HFP. At large time delays, $\theta
\geq1/\tilde{\gamma}$, both $p(\theta)$ and $q(\theta)$ are obviously slowly
decaying, with the decay being determined by the HFP decay. Thus, at large
times the difference $q^{2}-p\approx\tilde{\gamma}(p-p_{T})$. Considering
that the difference $q(\theta)^{2}-p(\theta)\ll p(\theta),q(\theta)$ and
that it is decaying slowly, an analytic approximate solution was derived 
\cite{RT}, describing the recovery dynamics: 
\begin{equation}
-2\tilde{\gamma}\theta=(2+\frac{1+2\tilde{\gamma}}{2q_{T}})\ln{[\frac{q-q_{T}%
}{q(\infty)-q_{T}}]}+(2-\frac{1+2\tilde{\gamma}}{2q_{T}})\ln{[\frac{q+q_{T}}{%
q(\infty)+q_{T}}]}  \label{qDecay}
\end{equation}
Here $q_{T}$ corresponds to the thermal QP density at base temperature and $%
q(\infty)$ is the value after the initial pre-bottleneck dynamics. Combining
solutions for pre-bottleneck case and recovery dynamics an approximate
solution for $q(\theta)$, valid over the entire time delay has been derived 
\cite{RT}, where 
\begin{equation}
q(\theta)=q_{1}(\theta)+q_{2}(\theta)-q(\infty),  \label{joint}
\end{equation}
with $q_{1}(\theta)$ given by Eq. (\ref{Q_RT}) and $q_{2}(\theta)$ by Eq.(%
\ref{qDecay}). The comparison of the analytic solution, given by Eq.(\ref%
{joint}), to the numerical solution of Eqs. (\ref{RTfullDim}), shown in
Figure 8, manifests an excellent agreement for all different limiting cases.

\begin{figure}[ptb]
\includegraphics[width=120mm]{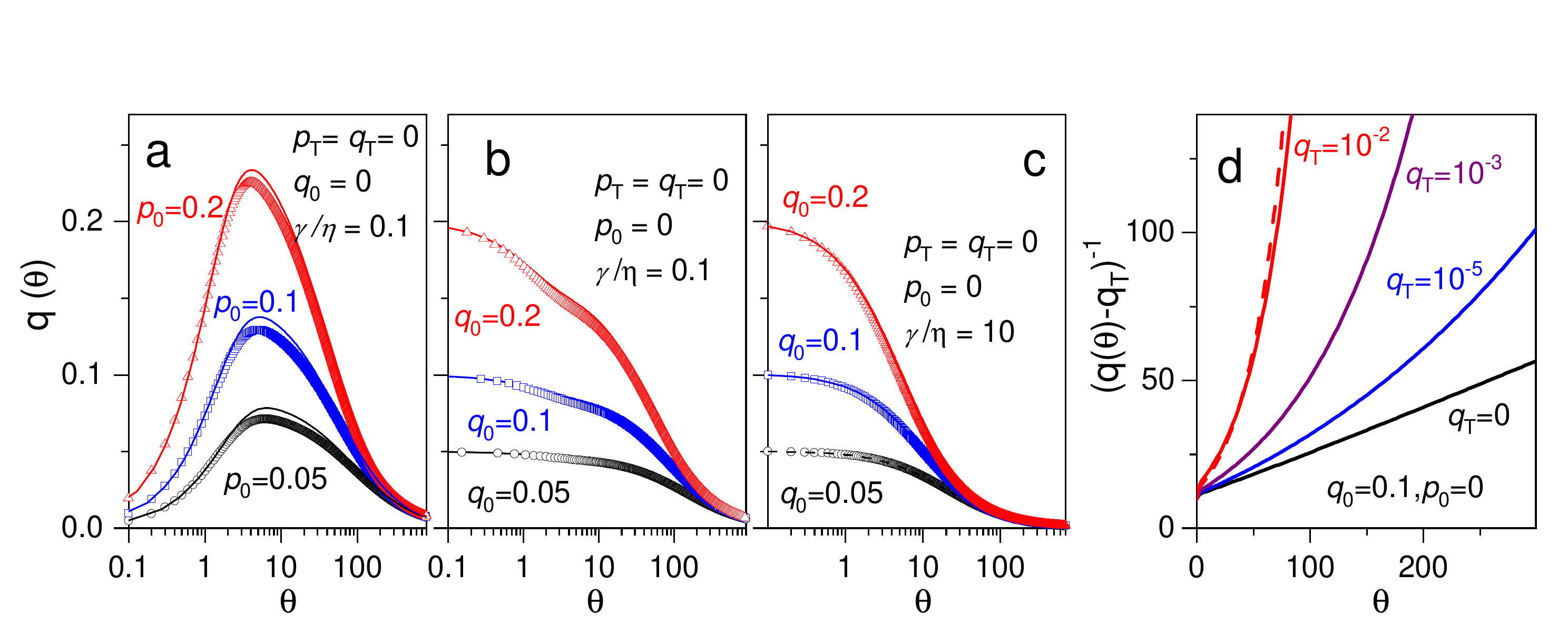} 
\caption{Analytic solutions of Eqs.(\protect\ref{joint}) (open symbols)
compared to numerical solutions (solid lines) in several limiting cases.
Panels (a) and (b) represent the strong bottleneck limit ($\tilde{\protect%
\gamma}\equiv\protect\gamma /\protect\eta=0.1$) for the two extreme
excitation conditions ($x=0$ and $1$, respectively) at $T=0$ K. Panel (c)
presents the case of weak bottleneck, i.e. when $\protect\gamma/\protect\eta%
=10$, at $T=0$ K. In all cases, the relaxation dynamics shows a strong
dependence of the SC recovery rate on excitation density. Finally, panel (d)
presents the dependence of the recovery dynamics on temperature, expressed
by the inverse of the photo-induced QP density as a function of $\protect%
\theta$ for various $q_{T}$. [reproduced from Ref. \protect\cite{RT}]}
\end{figure}

Figure 8 shows that, irrespective of the regime (strong or weak bottleneck),
the SC state recovery dynamics displays excitation density dependence at low
temperatures. Indeed, at $T\rightarrow0$ K, the time dependence of the
inverse of the photo-induced QP density is a linear function of $\theta$,
mimicking the pure bimolecular kinetics. At higher temperatures, the
relaxation becomes nearly exponential (dashed curve in Figure 8d presents
the exponential decay fit). We note that similar behavior was reported in
cuprates \cite{KaindlBiSCO}.

For the case of a strong bottleneck ($\gamma/\eta\equiv\tilde{\gamma}\ll1$),
the initial relaxation rate $\tau_{rec}^{-1}$ is given by \cite{RT} (the
expression here is valid for $k_{B}T\ll\Delta$) 
\begin{equation}
\tau_{rec}^{-1}=\frac{2\tilde{\gamma}(q(\infty)+q_{T})}{(1+2\tilde{\gamma})}.
\label{lowTtau}
\end{equation}
It is clear, that this expression qualitatively accounts for the
observations in NbN (Figure 7). In the limit of very weak excitations, where 
$q(\infty )\approx q_{T}$, the relaxation time should display exponential
temperature dependence with $\tau_{rec}^{-1}\propto q_{T}$. Indeed, this
accounts also the early studies of temperature dependence of the relaxation
time using biased tunnel junctions \cite{MillerDayem,Gray69}.

Moreover, the intensity dependence is observed if $q(\infty)\gg q_{T}$. In
this case $\tau_{rec}^{-1}$ is proportional to $q(\infty)$, which is in turn
proportional to excitation intensity (see Figure 5). It follows that the
excitation density dependence of the relaxation time should be observed at
low-temperatures, as demonstrated in cuprates \cite%
{KaindlBiSCO,GedikYBCO,PCCOBeck} as well as in conventional SCs like NbN 
\cite{Beck1}, and MgB$_{2}$ \cite{Sobolev}.

Further argument for the validity of the Rothwarf-Taylor description of the
light induced gap dynamics in superconductors follows from the comparison of
the experimental results on NbN (Figure 7) with the model predictions. The
data on NbN show, that the excitation and temperature dependence of $%
\tau_{rec}$ is observed for $\emph{A}\leq5$ mJ/cm$^{3}$ and for $T\leq10$ K.
For higher $\emph{A}$ and $T\geq10$ K $\tau_{rec}^{-1}$ is constant at $%
\tau_{rec}^{-1}(0)\approx0.01$ ps$^{-1}$, likely governed by the escape of
the HFP into the substrate (since it was found to be inversely proportional
to the film thickness \cite{Ilin}). In a phonon bottleneck case $%
\tau_{rec}^{-1}$ is proportional to the density of photo-excited QPs, $%
n_{PI} $, if $n_{PI}$ falls into the range $n_{T}<n_{PI}\leq\beta/R$. Here $%
\beta/R$ is the material dependent characteristic QP density, below which
the Rothwarf-Taylor perturbative description is applicable \cite{RT}.
Indeed, $A=5$ mJ/cm$^{3}$ corresponds to $n_{PI}\approx0.001$ unit cell$%
^{-1} $, identical to $\beta/R\approx0.001$ unit cell$^{-1}$ determined from
the analysis of the pair-breaking dynamics \cite{Beck1}. Similarly, the fact
that $\tau_{rec}^{-1}$ is constant above $\approx10$ K follows from the fact
that $n_{T}(10K)\approx0.0003$ ucv$^{-1}$ is comparable to $\beta/R$.
Therefore, as experimentally demonstrated for a standard BCS superconductor,
the \emph{A}-dependent SC state recovery at low temperatures seems to be an
expected behavior for a superconductor driven out of equilibrium.

We note that, while the phenomenological Rothwarf-Taylor has been derived to
describe the low-temperature dynamics in superconductors \cite%
{RothwarfTaylor,RT}, it was demonstrated to be able to qualitatively
describe also the dynamics in other systems with small energy gaps near the
Fermi level , i.e. when the gap is compared to the cut-off frequency of the
coupled bosonic excitation \cite{HFprl,HFreview}. E.g. in heavy fermion
metals a dramatic (exponential) slowing down of carrier relaxation was
observed at low temperatures \cite{HFprl,Ahn}, originally attributed to
slowing down of electron-phonon thermalization \cite{Ahn,HFPaper}.
Systematic studies on both heavy fermion metals and Kondo insulators
revealed that the dynamics can be accounted for by the presence of the
hybridization gap near the Fermi level, with the dynamics of QPs and coupled
HFP following the Rothwarf-Taylor bottleneck scenario \cite%
{HFreview,HFPaper,Rick,YbAl,Toni}.

\subsubsection{Superconducting state recovery dynamics in the limit $%
T\rightarrow T_{c}$}

In the early subsections we were focusing on the dynamics at low
temperatures, i.e. for $k_{B}T\ll\Delta$. Now we address the behavior near $%
T_{c}$ in the limit when $\Delta\rightarrow0$. 
\begin{figure}[ptb]
\includegraphics[width=120mm]{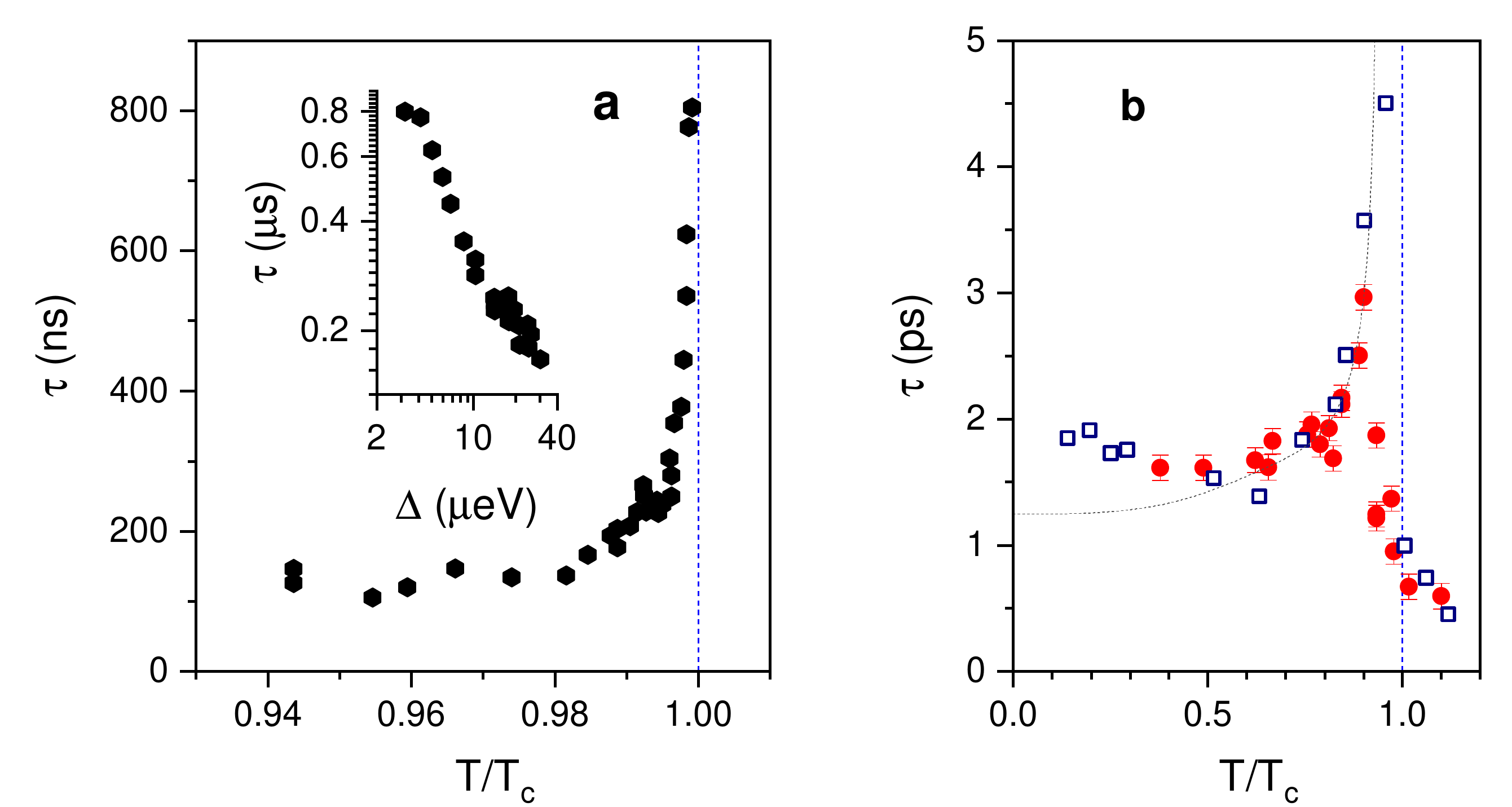} 
\caption{Dependence of the superconducting recovery timescale in the
vicinity of critical temperature. (a) Relaxation time in superconducting Al
as a function the reduced temperature \protect\cite{Schuller}. The
measurement were performed using optically excited tunnel junctions. Near T$%
_{c}$ the relaxation time shows 1/$\Delta$ dependence (inset). (b)
Relaxation time divergence near T$_{c}$ in photo-excited YBCO high
temperature superconductor. The open squares are from \protect\cite{Han},
the solid circles are from \protect\cite{DMM1998}. The dashed line is a fit
using a phonon bottleneck model, i.e. Eq.(\protect\ref{tau_critical}) 
\protect\cite{kabanov}. }
\end{figure}
As shown in Fig. 9a, already early experiments using real-time response of
the current--voltage characteristic of a tunnel junction to fast laser
pulses \cite{Schuller} revealed the anomalous slowing down of relaxation
upon approaching T$_{c}$. This can be interpreted by considering the
relaxation in terms of the Ginzburg-Landau theory of the second-order phase
transition. Here, the restoring force is given by the derivative of the free
energy with respect to the order parameter, giving rise to a divergence of $%
\tau_{rec}\propto$ $1/\Delta$. Considering the phonons being in thermal
equilibrium, Schmid and Schoen derived the expression to describe the
dependence of the superconducting state recovery time near T$_{c}$ \cite%
{SchmidSchoen}:%
\begin{equation}
\tau_{rec}=3.7\tau_{E}\frac{k_{B}T}{\Delta},
\end{equation}
where $\tau_{E}$ is the inelastic electron scattering time.

Similarly to early measurements using tunnel junctions, the divergence of
superconducting state recovery time was observed in superconductors using
optical pump-probe spectroscopy. Typical results obtained on YBa$_{2}$Cu$%
_{3} $O$_{7}$ are shown in Fig. 9b. Assuming the SC state recovery being
governed by the decay of HFPs, an alternative scenario, base on the
quasi-equilibrium between QPs and HFP was put forward \cite{kabanov}. Here,
Kabanov et al. \cite{kabanov} argued that the relaxation rate of the
photo-induced QPs is dominated by the energy transfer from HFP ($%
\hbar\omega>2\Delta$) to phonons with $\hbar\omega<2\Delta$, i.e. by an
anharmonic decay of HFP. Considering the kinetic equation for phonons,
taking into account phonon-phonon scattering (anharmonicity), the expression
describing the relaxation of the equilibrated QP-HFP temperature was derived:

\begin{equation}
\frac{1}{\tau_{rec}}=\frac{12\Gamma_{\omega}\Delta(T)^{2}}{%
\hbar\omega^{2}\ln\left\{ 1/(\emph{A}/2N(0)\Delta(0)^{2}+\exp(-%
\Delta(T)/k_{B}T))\right\} }.  \label{tau_critical}
\end{equation}
Here $\Gamma_{\omega}$ is anharmonic phonon decay rate given e.g. by the
phonon linewidth in Raman experiments, $N(0)$ is the normal state electronic
density of states, and $\emph{A}$ the absorbed energy density. Indeed, for
YBa$_{2}$Cu$_{3}$O$_{7}$ a good agreement of this formula with data (Figure
9b) is obtained.

Finally, we should mention another aspect of Kabanov's theory \cite{kabanov}%
, addressing the temperature dependence of the photo-induced QP density in
the vicinity of the critical temperature, when QPs and HFP are in
quasi-thermal equilibrium. Assuming that the energy gap is more or less
isotropic, the temperature dependence of the photo-excited QP density $%
n_{pe} $ (which corresponds to $\frac{\eta}{R}[q(\infty)-q_{T}]$ discussed
in section \ref{NbNBreaking}) is given by:

\begin{equation}
n_{pe}=\frac{\mathcal{A}/(\Delta(T)+k_{B}T/2)}{1+\frac{2\nu}{N(0)\hbar
\Omega_{c}}\sqrt{\frac{2k_{B}T}{\pi\mathbf{\Delta}(T)}}\exp(-\Delta
(T)/k_{B}T)},  \label{Kab_n_pe}
\end{equation}
where $\hbar\Omega_{c}$ is the phonon cut-off energy and $\nu$ the effective
number of phonon modes per unit cell (or phonon branches) participating in
the relaxation. Using this expression, and considering the BCS dependence of
the superconducting gap $\Delta(T)$, this expression is often used to
determine the gap magnitude even for systems, where $\Delta(T\rightarrow0)$
may not be spectroscopically determined. Examples of such analysis are
studies of the superconducting gap in cuprates \cite%
{DemsarYBCO,mercury,Dvorsek,KusarLSCO,Liu,Luo,TodaBisco} (see also section %
\ref{cuprates}), pnictides \cite{Mertelj,Chia,Kumar} and organic
superconductors \cite{Naito}. Furthermore, a similar expression \cite%
{kabanov} has also been derived for the case where the energy gap is
temperature independent, e.g. when discussing pseudo-gaps in cuprate
superconductors \cite{DemsarYBCO,mercury,Dvorsek,KusarLSCO,Liu,Luo,TodaBisco}%
.

\subsection{Superconducting gap dynamics in high-T$_{c}$ cuprate
superconductors}

\label{cuprates}

In the above sections we focused on the real-time dynamics in conventional
BCS superconductors with s-wave (isotropic) gap. Cuprate high-T$_{c}$
superconductors, however, are characterized by an anisotropic d-wave gap
with nodes (directions in the momentum space with vanishing gap). Thus, one
may expect their behavior under non-equilibrium conditions may be
substantially different. Surprisingly, however, there seem to be more
similarities than differences between the two classes of superconductors in
terms of dynamics. Instead of a detailed review of the field of dynamics in
cuprate superconductors, which can be found in Ref. \cite{Giannetti}, we
briefly review the similarities and anomalies with respect to the
observations in conventional BCS s-wave superconductors (discussed in
sections above).

\begin{figure}[ptb]
\includegraphics[width=120mm]{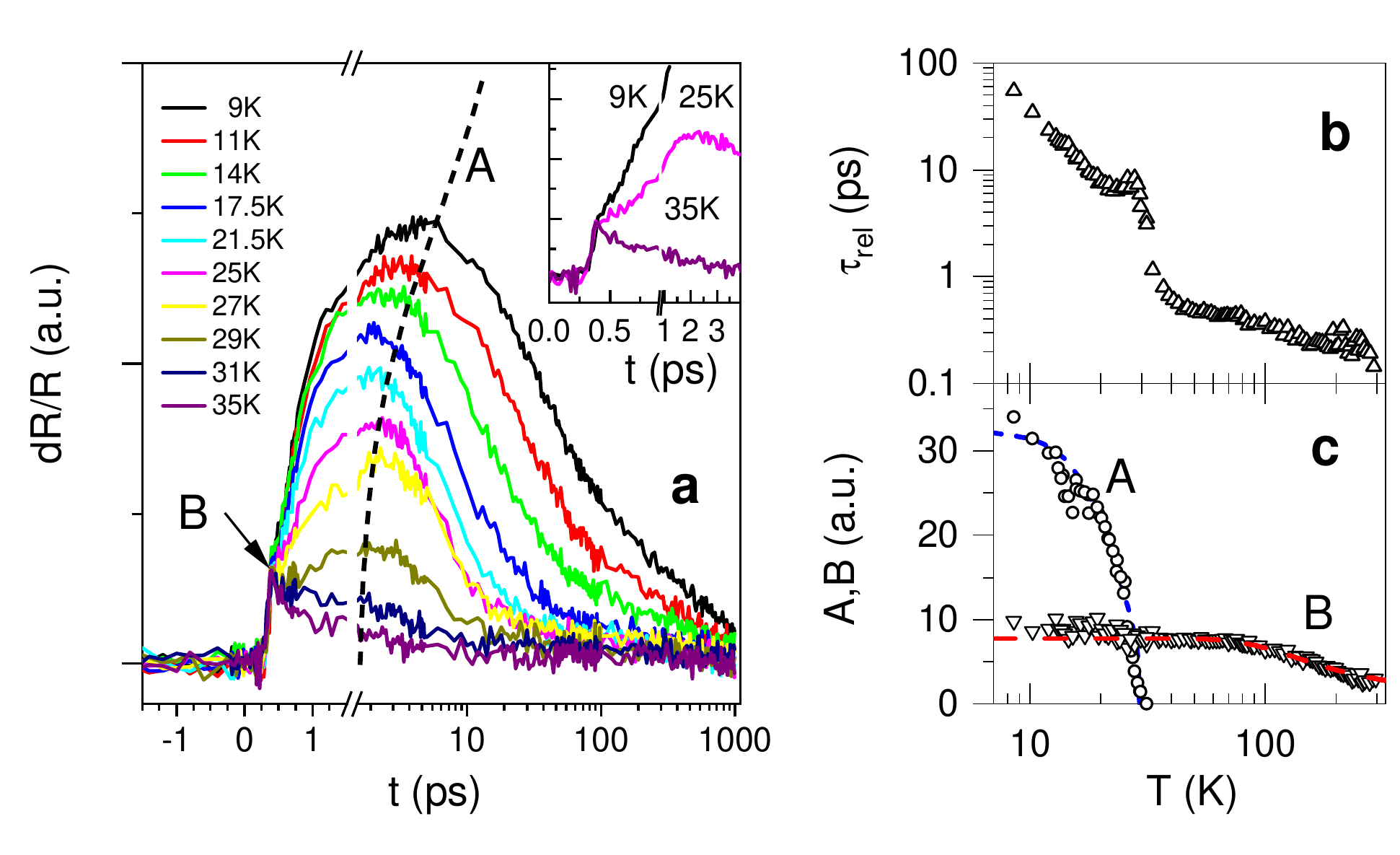} 
\caption{(a) Dynamics of photo-induced change in reflectivity $\Delta$R/R in
La$_{1.9}$Sr$_{0.1}$CuO$_{4}$ single crystal (T$_{c}=30$ K) recorded in the
low excitation density limit. Two distinct components (A and B) are observed
when probing dynamics with near-infrared pulses (see also inset). (b) The
temperature dependence of relaxation time of the reflectivity signal. (c)
The temperature dependence of amplitudes of the two components, including
fits with the model by Kabanov et al. \protect\cite{kabanov}. For details,
see the original publication \protect\cite{KusarLSCO}, from which the data
are reproduced. }
\end{figure}

Figure 10a presents the typical data obtained on cuprate superconductors in
the low excitation density limit. It shows the temperature dependence of
photo-induced reflectivity ($\Delta R/R$) traces on La$_{1.9}$Sr$_{0.1}$CuO$%
_{4}$ single crystal (T$_{c}=30$ K) \cite{KusarLSCO}. The data are often
described in terms of a two component response, with component A (see Figure
10a) present only below T$_{c}$. Component B, however, is present to
temperatures far above T$_{c}$. Probe polarization \cite{Dvorsek,TodaBisco}
and chemical doping\ \cite{DemsarYBCO,KusarLSCO} dependent studies revealed
the coexistence of the two components below T$_{c}$. On the other hand, the
correlation of the (doping-dependent) onset temperature of the component B 
\cite{DemsarYBCO,KusarLSCO} has been made with the so called "pseudo-gap
temperatures" determined by other methods, sensitive to the changes in the
low energy density of states \cite{Zagar}. These suggest the origin of the
component B being in the response of the competing order, like e.g.
(fluctuating) charge stripes or density waves.

From here on, we focus on the component (A) that can be clearly linked to
the dynamics of the superconducting condensate, as demonstrated in Figure
11. Here, a comparison of the dynamics, recorded by (a) probing the changes
in reflectivity at 800 nm and (b) the THz conductivity as a function of
excitation density (T = 4 K) is presented. At low excitation densities,
where the fast relaxing component (B) associated to the dynamics of the
competing order is small compared to the response of the condensate, both
responses (reflectivity changes and changes in the THz conductivity) are
comparable. Only for high excitation densities, when superconductivity
becomes strongly suppressed, i.e. when the photoinduced change is
saturating, the difference between the two responses becomes obvious. We
note that the fast relaxing component B is absent (or dramatically reduced)
in the THz response. The reason for that lies in the nature of the THz
response of a superconductor, which is governed by the change of the phase
of the transmitted THz electric field pulse (caused by the inductive
character of the complex optical conductivity in the superconducting state)
- see inset to Figure 11b.

\begin{figure}[ptb]
\includegraphics[width=120mm]{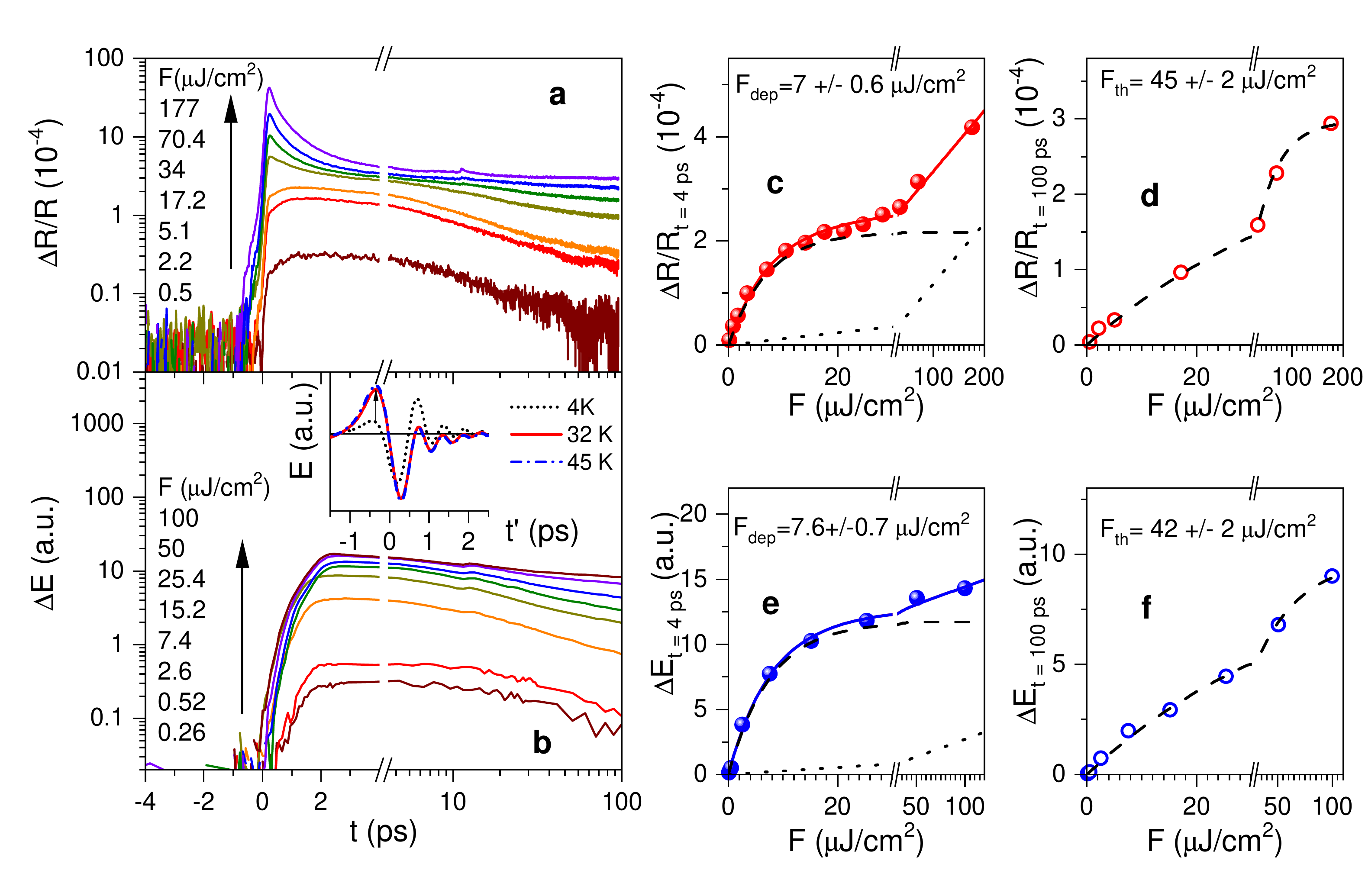} 
\caption{Excitation density dependence of dynamics in La$_{1.84}$Sr$_{0.16}$%
CuO$_{4}$ thin film (T$_{c}=31$ K) at 4 K, excited by 50 fs near-infrared
(800 nm) pulses. The dynamics are probed in two different experimental
configurations: (a) by recording changes in reflectivity at 800 nm and (b)
by probing changes in THz conductivity (spectrally integrated approach).
Panels (c) and (d) present the excitation dependence of reflectivity change
at (c) 4 ps and (d) 100 ps after photo-excitation. The corresponding
amplitudes of the change in the electric field at a fixed time delay (see
inset to panel (b)) at the two distinct time delays are shown in panels (e)
and (f), respectively. Note that the THz electric field is mainly sensitive
to superconductivity, as exemplified in inset to panel (b). At long time
delays (100 ps) the system is already thermalized, and the corresponding
response is a bolometric one. Indeed, the values for characteristic fluence
require to heat up the film to above T$_{c}$, F$_{th}$, perfectly match the
thermodynamic calculation \protect\cite{Beyer}. In panels (d) and (f) The
presented data are taken from Ref. \protect\cite{Beyer}. }
\label{LSCOnir}
\end{figure}
Now, let us address the similarities (and some surprising inconsistencies)
of the condensate dynamics in high-T$_{c}$ cuprates with conventional
superconductors.

\subsubsection{Pair-breaking dynamics in cuprates}

Looking at Figures 10a, 11a and 11b, one realizes a similar delayed
pair-breaking dynamics as in NbN and MgB$_{2}$ shown in Figure 4. While the
timescale is substantially shorter (depends also on the compound), one can
argue in favor of pair-breaking by absorption of bosonic excitations created
during the avalanche relaxation of hot QPs. This result suggests the same
type of boson-bottleneck governing the relaxation as in conventional
superconductors. Here, considering the possible scenarios of
superconductivity in cuprates being mediated by phonons or magnetic
excitations, we cannot specify a priori which bosonic excitation may be
responsible for pair-breaking (or pairing).

We should further mention the study on YBa$_{2}$Cu$_{3}$O$_{7}$ crystals,
where the dynamics of both, the condensate and specific IR-active c-axis
phonons were studied using time-resolved broadband THz conductivity \cite%
{Pashkin}. The dynamics in the superconducting state is presented in Figure
12. Panel (a) presents the in-plane response, showing a rapid filling-in of
the superconducting gap (equilibrium $\sigma_{1}(\omega)$ in the
superconducting and normal state are shows at the top) and its recovery on
the timescale of a few ps. The dynamics is similar to the results of earlier
studies using near-infrared probe \cite{DemsarYBCO} of 0.2-1 THz pulses \cite%
{THzAveritt}. Panel (b) presents the out-of-plane (c-axis) response,
involving the dynamics of two c-axis polarized phonons. Especially the apex
oxygen mode at $\approx17$ THz is found to be strongly affected by
superconductivity, giving rise to a highly asymmetric line-shape in the
superconducting state. The analysis of the time-evolution of the phonon
line-shape in panel (c) provides access to the dynamics of the condensate,
related to the phonon asymmetry parameter shown in (d), and the phonon
frequency. The later, presented in panel (e), displays a rapid red-shift of
the mode and its sub-picosecond recovery. Since the red-shift is
characteristic of an anharmonic lattice potential, the shift of the apex
phonon frequency can be used as a probe of the vibrational occupation.
Indeed, the presented frequency shift in Figure 12e suggest a rapid heating
of the mode, reaching occupations compared to those achieved by heating up
to 200 K \cite{Pashkin}. The results indicate highly excited lattice modes
on the timescale comparable to pair-breaking.

\begin{figure}[ptb]
\includegraphics[width=120mm]{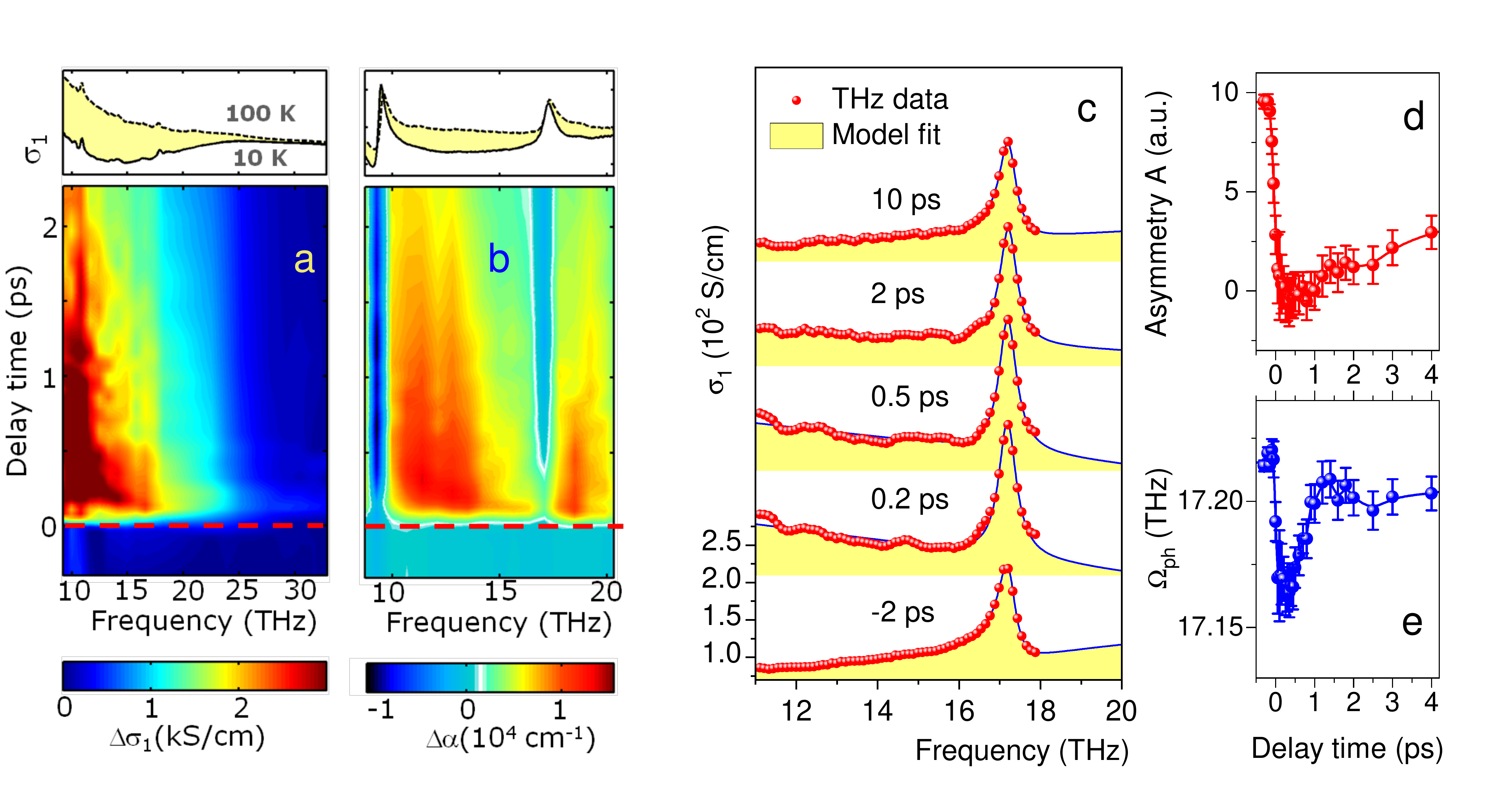} 
\caption{Spectrally resolved studies of the broadband THz conductivity in YBa%
$_{2}$Cu$_{3}$O$_{7}$ at 20 K\protect\cite{Pashkin}. (a) In-plane response,
showing photo-induced filling of the gap and its subsequent recovery on the
ps timescale. (b) c-axis response, capturing the dynamics of 2 c-axis
phonons. (c) The line-shape analysis of the apex oxygen mode at 17.2 THz 
\protect\cite{Pashkin}. The time-evolution of the two relevant parameters,
the asymmetry parameter (d) and the mode central frequency (e). Data are
reproduced from Ref. \protect\cite{Pashkin}.}
\end{figure}

\subsubsection{Superconducting state recovery dynamics in cuprates}

The relaxation time in many of the cuprate superconductors shows similar
temperature \cite{mercury,GedikYBCO,KusarLSCO,Schneider} (see also Figure
10b) and excitation density dependence \cite{KaindlBiSCO,PCCOBeck,GedikYBCO}
as observed in NbN \cite{Beck1} and predicted by the Rothwarf-Taylor model
for both weak and strong bottleneck cases \cite{RT}. Together with the
observed delayed pair-breaking, discussed above, the data seem to be
generally consistent with the strong boson-bottleneck scenario. On the other
hand, some doping dependent studies of carrier relaxation dynamics suggest
that the character of relaxation may change between the under-doped and the
over-doped regions of the phase diagram. In particular, Gedik et al. \cite%
{GedikBiSCOdoping} observed that the relaxation time is temperature and
excitation density dependent in the under-doped case, while in over-doped at
low temperatures the gap recovery dynamics is independent on excitation
density and temperature. Indeed, the observation in Bi$_{2}$Sr$_{2}$Ca$%
_{1-y} $Dy$_{y}$Cu$_{2}$O$_{8}$ series \cite{GedikBiSCOdoping} is consistent
with the observation on YBa$_{2}$Cu$_{3}$O$_{7-\delta}$ samples, where in
the over-doped range the relaxation dynamics at low temperatures seems to be
independent on temperature and excitation density \cite{DemsarYBCO}. On the
other hand, systematic doping dependence studies on single-layer cuprates do
not seem to show this disparity \cite{KusarLSCO,Schneider}.

\subsubsection{Energetics of the gap suppression in cuprate superconductors}

There are two important aspects of dynamics in cuprates that are revealed by
studying excitation density dependent suppression of superconductivity.

First of all, in the low excitation density limit, the photo-induced changes
in the QP density in cuprates scale linearly with the excitation density.
Such an observation is straightforward for the case of a superconductor with
an isotropic gap. Here the number of broken pairs per an absorbed photon
with energy $\hbar\omega$ is simply given by $n_{pe}=\hbar\omega/2\Delta$
(or a fraction thereof - see section \ref{NbNBreaking}). But, for an
anisotropic gap with nodes this observation is unexpected \cite{kabanov}.
Here, the quasiparticle density of states for energies $\varepsilon\ll%
\Delta_{a}$, where $\Delta_{a}$ is the maximal gap in the anti-nodal
direction, is given by $N(\varepsilon)=N(0)\left( \frac{\varepsilon}{%
\Delta_{a}}\right) ^{\kappa}$ (power $\kappa$ depends on the topology of the
nodes on the Fermi surface). Therefore, at low temperatures, a sub-linear
behavior of the photo-induced QP density with excitation intensity is
expected \cite{kabanov}, with $n_{pe}\propto\emph{A}^{(\kappa+1)/(\kappa+2)}$%
. For quasi 2D cuprates, with $\kappa=1$, this would suggest $n_{pe}\propto$ 
$\emph{A}^{2/3}$ dependence at low temperatures, which was never observed.

In fact, even the temperature dependence of $n_{pe}$ extracted from
time-resolved experiments, seems to be at odds with the behavior expected
for a d-wave superconductor \cite{kabanov,nicol}. A possible origin of this
apparent contradiction may be in a slow relaxation of hot QPs towards the
nodes due to kinematic constraints (i.e. the energy and momentum
conservation) \cite{Hirschfeld}.

Another interesting aspect related to the excitation dependent\ studies of
suppression of superconductivity in cuprates is related to the energy
required to deplete superconductivity. As shown in section \ref{NbNEnerg},
the absorbed energy density required to suppress superconductivity roughly
matches the condensation energy in NbN \cite{Beck1}. Systematic studies on
cuprates \cite{Kusar,Beyer,PCCOBeck}, however show that the energy required
to suppress superconductivity exceeds the condensation energy by about one
order of magnitude. Figure 11 present the fluence dependence of the
superconducting state depletion in La$_{1.84}$Sr$_{0.16}$CuO$_{4}$ using
near-infrared pump pulses in two configurations. The saturation of the
photo-induced change in reflectivity or THz conductivity at 4 ps time delay
is observed at the fluence $\approx7$ $\mu$J/cm$^{2}$, which corresponds to
the absorbed energy density of $2.4$ k$_{B}$K per Cu atom. This energy is
about 8 times higher than the condensation energy $\approx0.3$ K/Cu \cite%
{CpLSCO}. One possible reason for this discrepancy is a large magnitude of
the superconducting gap in La$_{2-x}$Sr$_{x}$CuO$_{4}$, compared to the
Debye energy. In other words,\ in La$_{2-x}$Sr$_{x}$CuO$_{4}$ there are
several optical phonon branches with $\hbar\omega$ $<2\Delta_{a}$ ($%
\Delta_{a}$ is the gap energy in anti-nodal direction), thus a large amount
of the absorbed optical energy can be transferred to phonons with $%
\hbar\omega$ $<2\Delta_{a}$, on the 100 fs timescale. Since pair-breaking by
phonons with $\hbar\omega$ $<2\Delta_{a}$ is strongly suppressed, these
could serve as an energy sink during the initial avalanche relaxation of hot
carriers towards the gap edge.

Recently, similar study in the electron-doped cuprate superconductor Pr$%
_{2-x}$Ce$_{x}$CuO$_{4}$ was performed \cite{PCCOBeck}. Here melting of the
superconducting state with both near-infrared and THz pulses was studied.
Using THz pulses, tuned just above the gap, the absorbed energy density
required to deplete SC was found to match the condensation energy. With
near-infrared excitation, however, the absorbed energy density was again a
factor of 5-6 higher than the condensation energy - despite the fact that in
Pr$_{2-x}$Ce$_{x}$CuO$_{4}$ the gap, $2\Delta $, is small compared to all
relevant bosonic energy scales \cite{PCCOBeck}. The data imply that
following optical pumping a rapid \textit{electron-boson} energy transfer
takes place, yet only selected bosonic modes (e.g. antiferromagnetic
fluctuations, or specific lattice modes) couple to the condensate. Further
systematic studies, with selective (tuned) pumping are clearly required here.

\subsubsection{Open questions}

Most of the observations obtained by systematic temperature and excitation
dependent studies of the dynamics in the superconducting state show great
similaritis between cuprates superconductors and conventional BCS
superconductors. Both pair-breaking and the superconducting state recovery
dynamics are found to be consistent with a boson-mediated pairing in
cuprates. The nature of the pairing "glue" is, however, still an open
question. On the other hand, unlike in BCS superconductors, photoexcitation
does not seem to give rise to a reduction of the gap magnitue (as seen for
NbN in Fig. 3), but rather to filling-up of the gap, i.e. to an increase of
the spectral weight inside the gap \cite{Dessau}.

\section{Enhancement of superconductivity by resonant excitation}

\subsection{THz driven ehnancement of superconductivity in BCS
superconductors}

In section \ref{EnhancHistory} we addressed the early studies of enhancement
of superconductivity with sub-gap excitation. Most of these experiments were
conducted using continuous excitation with sub-gap radiation, and the
effects (increase of the superconducting gap, critical current and
transition temperature) were typically of the order of 1 \%, limited by the
QP scattering rate and the radiation induced sample heating. Using the novel
pulsed sources of intense infrared radiation one may expect to observe
transient effects, whose magnitude would exceed magnitudes observed in early
studies. Here we briefly review a study, which suggest superconductivity
enhancement effects in a dirty-limit superconductor NbN \cite{Beck13} on a $%
100$ ps timescale, when the system is driven by narrow-band THz pulses tuned
close to the gap energy.

\begin{figure}[ptb]
\includegraphics[width=120mm]{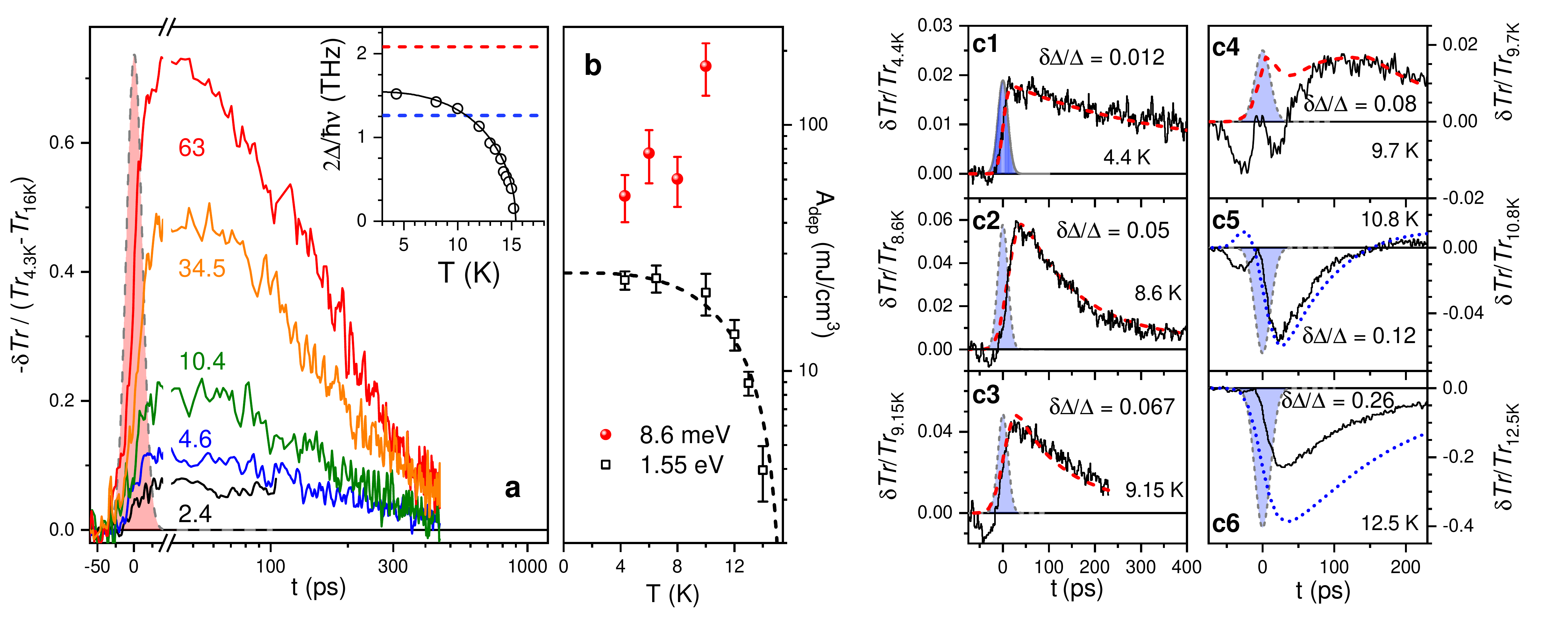} 
\caption{Dynamics in NbN films driven by narrow-band THz excitation 
\protect\cite{Beck13}. (a) Dynamics of the relative change in transmission $%
\protect\delta $\textit{Tr} (normalized to \textit{Tr}$_{4.3K}$-\textit{Tr}$%
_{16K}$) at 4.3 K when pumping with 2.08 THz pulses of different absorbed
energy density $\emph{A}$ (in mJ/cm$^{3}$). The shaded region presents the
pump-probe intensity auto-correlation. Inset shows the T-dependence of gap
with the two frequencies used in narrow band THz pump-probe experiments by
the dashed lines. (b) Comparison of the absorbed energy density required to
deplete superconductivity, measured with near-infrared pumping at 1.55 eV 
\protect\cite{Beck1} and THz pumping at 8.3 meV (2.08 THz). The dashed line
represents the condensation energy. (c1-c6) Recorded traces of differential
transmission for pumping at $1.26$ THz at different temperatures (black
solid lines). The shaded region presents the pump-probe intensity
auto-correlation. For red dashed lines present the behavior in the case of
the so called T$^{\ast}$-model, which adequately described the dynamics for
near-infrared excitation \protect\cite{Beck1}. At each temperature the
maximum relative change in the gap $\protect\delta\Delta/\Delta$ is
estimated from the measured $\protect\delta$\textit{Tr}/\textit{Tr}.}
\end{figure}

In the study, summarized in Figure 13, narrow band THz pulses ($\approx30$
GHz bandwidth, 20 ps duration) were used to both excite NbN and probe the
resulting changes in the transmission. The studies using near-infrared
excitation \cite{Beck1} demonstrated, that the time evolution of the NbN
superconductor can be, for time delays beyond the pair-breaking process,
described by the time evolution of effective temperature T* of thermalized
condensate, QPs and HFPs. As the transmission through the sample is strongly
affected by the opening of the superconducting gap, one has - in principle -
access to the time evolution of the gap, without\ having spectral
resolution. With these assumption, two types of experiments were performed:
above gap pumping at $2.08$ THz, with traces shown in panel (a), and pumping
with $1.26$ THz pulses (which is below 2$\Delta$ for T $< 11$ K - see inset
to Figure 13a) with traces at different temperatures shown in panels
(c1)-(c6).

Two interesting observations were made, both suggesting an additional
mechanism present at temperatures close to T$_{c}$, competing with the
expected light induced suppression of superconductivity.

In experiments with above-gap excitation, the absorbed energy density
required to deplete superconductivity, \emph{A}$_{\text{dep}}$, was
determined at different base temperatures, similar to experiments performed
with near-infrared pumping (see Figure 6). The surprising observation is
that \emph{A}$_{\text{dep}}$\emph{\ }is increasing with increasing
temperature (Figure 13b) opposite to the experiments with near-infrared
(1.55 eV) pumping. This is a first indication that an additional process may
be competing with the common light-driven suppression of superconductivity.

Even stronger evidence of a competing process is provided by pumping at 1.26
THz. Panels (c1)-(c6) present the recorded traces of the relative
transmission change, compared to the simulations assuming the T$^{\ast}$
model, i.e. the assumption that the superconducting state at $\sim30$ ps
after photo-excitation can be described by $\Delta^{\ast}=\Delta($T$%
^{\ast}). $ The simulations are given by the dashed red and dotted blue
lines. The non-monotonous behavior is a result of the fact, that at\ 1.26
THz the transmission is a non-monotonous function of temperature \cite%
{Beck13}. For time delays less than $\approx50$ ps, the relative
transmission changes systematically shows the opposite direction to the one
predicted by the T$^{\ast}$-model. This is especially pronounced for
near-resonant pumping, when $2\Delta/h\approx1.26$ THz, around 10 K. This
observation suggest a transient gap enhancement (and the corresponding
increase in condensate density) under THz pumping for timescales less than $%
\approx50$ ps.

Given the fact that these effects are pronounced in the vicinity of critical
temperature, when the density of thermally excited QPs becomes prominent,
the data are interpreted within the Eliashberg scenario \cite%
{Eliashberg70,Beck13}. Here, narrow-band excitation gives rise to a highly
non-thermal QP distribution resulting in amplification of SC that would be
washed out under continuous excitation. Taking the data at face value, the
enhancement effects seem to be observed even in the case of above gap
excitation, where QP excitation competes with pair-breaking. One could
imagine that such experiments, if performed with pulses at even lower
carrier frequencies and closer to the critical temperature, may even lead to
stronger effects as discussed above.

\subsubsection{Photo-induced superconductivity in unconventional
superconductors}

\label{photoinduced_unconventional}

Following the ideas of the pioneering works on radiation enhances
superconductivity addressed in section \ref{EnhancHistory}, we discussed
first experiments utilizing intense narrow band pulses, aiming to
transiently enhance superconductivity in conventional BCS superconductors.
We should, however, also briefly address more recent studies reporting
superconducting-like transient states (with the typical duration of 1 ps)
observed in cuprate superconductors \cite{Fausti,Wu,Mankowsky,Kaiser},
fullerides \cite{Mitrano,NatPhys,Cantaluppi,Budden} and recently also on
organic charge-transfer salts \cite{Organics} when photoexciting at base
temperatures far above the corresponding equilibrium critical temperatures
in these systems. The leading idea in these experiments is to excite
specific lattice modes/distortions by utilizing non-linear phononics \cite%
{Foerst,CavalleriRev}. This way, pulsed excitation can either suppress a
competing (stripe) order \cite{Fausti,Averitt}, or induce superconductivity
by the lattice-driven modification of the pairing interaction \cite%
{Fausti,Wu,Mankowsky,Kaiser,Mitrano}.

\begin{figure}[ptb]
\includegraphics[width=100mm]{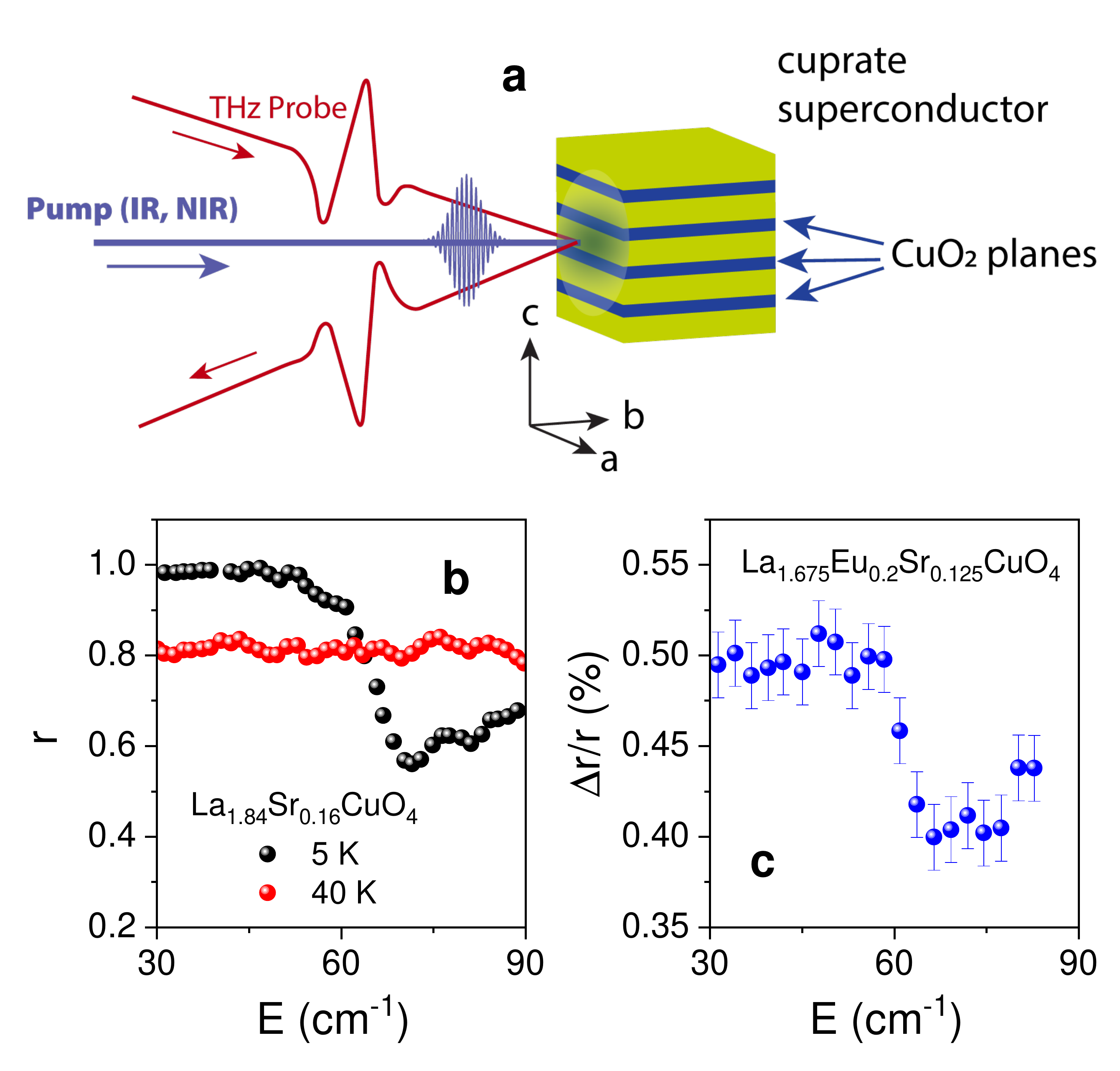} 
\caption{Studies of photoinduced superconductivity in cuprate
superconductors. (a) A typical experimental configuration, probing the
induced changes in the out-of-plane (c-axis) THz conductivity following
mid-infrared (or near-infrared) excitation. (b) The changes in equilibrium
c-axis reflectance ($r=E_{refl}/E_{inc}$) upon cooling the optimally doped
LSCO (La$_{1.84}$Sr$_{0.16}$CuO$_{4}$) through T$_{c}=38$ K (adapted from
Ref. \protect\cite{Fausti}). (c) The photoinduced change in reflectance in
stripe-ordered (non-superconducting) La$_{1.675}$Eu$_{0.2}$Sr$_{0.125}$CuO$%
_{4}$ after photoexcitation with infrared pulse at 16 $\protect\mu$m
(adapted from Ref. \protect\cite{Fausti}) While the overall reflectance
change is positive in the THz spectral range, the part of the response
having a step-like shape was attributed to the light induced
superconductivity.}
\label{LESCO}
\end{figure}

Figure \ref{LESCO} presents the experimental approach and the results
obtained in one of the first studies of this type, the study of
light-induced melting of the competing stripe order in La$_{1.675}$Eu$_{0.2}$%
Sr$_{0.125}$CuO$_{4}$ (LESCO$_{1/8}$) \cite{Fausti}. The non-superconducting
LESCO$_{1/8}$ is known to posess a static stripe order. Upon excitation with
infrared pulses, the induced changes in reflectance (Fig. \ref{LESCO}c)
displays a shoulder-like structure around 60 cm$^{-1}$, which was
interpreted as a photoinduced c-axis Josephson plasma resonance (JPR) \cite%
{Fausti}. The JPR is a general feature of layered cuprate superconductors,
explained by the Josephson coupling between the adjecant superconducting (CuO%
$_{2}$) layers, and can be observed in equilibrium studies of c-axis optical
conductivity, as shown in Fig. \ref{LESCO} b. The fact that the induced
changes in reflectance, $\Delta r/r$, are small (the step in $\Delta r/r$,
attributed to the JPR is only about 0.1\%, to be compared to the $\approx 40$%
\% effect in the equilibrium), could be attributed to a large mismach of the
penetration depths. Indeed, the pump field penetration depth at 16 $\mu $m
is about 200 nm while that of the THz probe field is about 10 $\mu $m \cite%
{Fausti}. The comaparably large, frequency independent offset in $\Delta r/r$
was however attributed to the photoinduced changes in the high(er) frequency
phonons \cite{Fausti}. By processing the raw data, taking into account the
large penetration depth mismatch and subtracting a frequency independent
component originating from perturbed higher-frequency modes, the authors
showed that $\sigma _{2}$ from the photoxcited layer displayed the upturn at
low frequencies, resembling the response of the superconductor \cite{Fausti}%
. The fact that the effect was peaked around the pump-photon energy of 80
meV (15 $\mu $m) lead to the interpretation of photoinduced
superconductivity via melting of the underlying static stripe order \cite%
{Fausti}. The result is particularly surprising, considering the\ excitation
densities in these studies are of the order of 1 mJ/cm$^{2}$, nearly two
orders of magnitude higher than excitation densities required to suppress
superconductivity in an optimally doped LSCO, when excited by NIR pulses
polarized in the $a-b$ plane (see Fig. \ref{LSCOnir}).

Following this pioneering work, the approach has been extended to the
superconducting cuprates. Using the same approach, it was shown that the
transient state with the low-frequency divergence in $\sigma _{2}\left(
\omega \right) $, reminescent of $1/\omega $ inductive response of the
condensate, could be realized over large range of doping and temperatures,
approaching room temperature in strongly underdoped cuprates \cite%
{Wu,Kaiser,CavalleriRev}. The interpretation of these data is, however,
still under intense discussion, ranging from aspects of the data analysis to
the different interpretations of the origin of the light-induced
modifications of low frequency $\sigma \left( \omega \right) $ \cite%
{Dodge,Response,Shimano,NLWang,Demler,Manske,Nava,Millis,Millis2020}. Recently,
several groups performed similar studies of the c-axis THz response using
both mid-infrared as well as near-infrared pulses for photoexcitation \cite%
{Averitt,Shimano,NLWang,Wang2018,Wang2018a,Liu2019}. While it was
demonstrated that near-infrared excitation results in similar changes in the
low frequency conductivity as with mid-infrared excitation \cite%
{Wang2018,Liu2019}, the consensus on the nature of the photoinduced state is
still lacking. In the superconducting state of the underdoped LSCO Niwa et
al. \cite{Shimano} observed a photoinduced red shift of the JPR, consistent
with the photoinduced suppression of superconductivity. It was argued that
transient heating linked to the inhomogeneous excitation profile could give
rise to potential artefacts in the analysis \cite{Shimano}. Studies in the
superconducting state of La$_{1.905}$Ba$_{0.095}$CuO$_{4}$ \cite{Wang2018},
La$_{1.885}$Ba$_{0.115}$CuO$_{4}$ \cite{Averitt}, and Pr$_{0.88}$LaCe$_{0.12}
$CuO$_{4}$ \cite{Wang2018a}, however reported a photoinduced red shift of
the JPR and the appearence of an additional mode at higher frequencies.
While at temperatures above T$_{c}$ no such photoinduced mode was observed
in La$_{1.905}$Ba$_{0.095}$CuO$_{4}$ \cite{Wang2018}, it has been observed
in La$_{1.885}$Ba$_{0.115}$CuO$_{4}$, albeit at very high excitation density
of 9 mJ/cm$^{2}$ \cite{Averitt}. Finally, following up on the first reports,
suggesting signatures of transient (inhomogeneous) superconductivity up to
room temperature in underdoped YBa$_{2}$Cu$_{3}$O$_{6+x}$ \cite{Wu,Kaiser},
similar studies have been performed by both mid-infrared and near-infrared
excitation \cite{NLWang}. While, similarly to earlier reports \cite%
{Wu,Kaiser}, a photoinduced increase of THz reflectivity was observed, the
authors argue against the photoinduced (inhomogeneous) superconductivity 
\cite{NLWang}. From the above, it is clear that further systematic studies
are required. Ideally, the experiments should be performed on a-axis
oriented thin films in transmission geometry, to insure homogeneous
excitation profile, and to enhance the experimenal sensitivity (as shown in
section \ref{realTimeSpectroscopy}, superconductivity results in vehement
changes in thin film transmission).

We note that similar studies to those in cuprates were performed also on K$%
_{3}$C$_{60}$ \cite{Mitrano,NatPhys,Cantaluppi,Budden} and recently also on
organic charge-transfer salt $\kappa $-(BEDT-TTF)$_{2}$Cu[N(CN)$_{2}$]Br 
\cite{Organics}. In both materials, upon photoexcitation with intense
mid-infrared pulses, spectral signatures of superconductivity were reported
for base temperatures well above the corresponding equilibrium critical
temperatures. In K$_{3}$C$_{60}$ the mismatch between the pump and the probe
penetration depths was still large \cite{Mitrano}, requiring multilayer
modelling (as in cuprates). This does not seem to be the case for the
organic charge-transfer salt \cite{Organics}. At face value, the extracted
optical conductivities in the transient states indeed seem like to result
from photoinduced superconductivity, i.e. showing gapping (reduction) of $%
\sigma _{1}$ and $\sigma _{2}\left( \omega \right) \propto 1/\omega $. In K$%
_{3}$C$_{60}$, where the recent study suggests the photoinduced state to be
long-lived\ with lifetime well in the nanosecond range \cite{Budden}, the
optical conductivity data is supplemented by the DC transport studies,
demonstrating the photoinduced drop in resistivity \cite{Budden}. 

Despite the apparent agreement with the proposed scenario of
light-enhanced/induced superconductivity, where mid-infrared pulses induce
large-amplitude structural distortions and thereby modulate the pairing
strength \cite{Mitrano,Budden}, there are still several open questions.
E.g., experiments on K$_{3}$C$_{60}$ performed at temperatures below T$_{c}$
show a photoinduced suppression of superconductivity \cite{Mitrano} instead
of the expected gap enhancement. Apart from the proposed exciton-cooling
model \cite{Nava} this observation seems at odds with majority of the
proposed scenarios \cite{Mitrano,Demler,Kennes,Kim17,Mazza}. Secondly, the
excitation densities in most of these experiments are extremely high. E.g.,
the recent study reporting the generation of a long-lived (nanoseconds)
metastable superconducting state in K$_{3}$C$_{60}$ \cite{Budden} suggests a
fluence threshold for the effect of about 20 mJ/cm$^{2}$. With the pump
penetration depth of 200 nm, and the reflectivity of 30 \% \cite%
{Mitrano,Budden} one obtains the absorbed energy density of the order of 700
J/cm$^{3}$. The resulting temperature of the excited volume, assuming the
system is quasi-thermalized on the 1 ns timescale, can be estimated from the
temperature dependence of the total specific heat of K$_{3}$C$_{60}$ \cite%
{CpK3C60}, and amounts to about 500 K! Here we should mention a recently
proposed alternative scenario,\ where the experimental data are accounted
for by considering that the pulse, resonant with a phonon mode, creates a
(nonsuperconducting) nonequilibrium state in which the linear response
conductivity becomes negative \cite{Millis,Millis2020}.

\section{Summary and perspectives}

In this brief review, we addressed some aspect of superconductors, driven
out of equilibrium using femtosecond and picosecond optical pulses. The
real-time experimental approaches and theoretical models presented here,
building on the pioneering works from the early era of non-equilibrium
superconductivity (sections \ref{DynHistory},\ref{EnhancHistory}), provide
new insights into this fascinating field of condensed matter physics.

This review is, however, far from comprehensive in view of addressing all
aspects related to light-driven out-of-equilibrium phenomena in
superconductors. We briefly discussed novel approaches, aiming to
impulsively modulate the pairing strength and thereby enhance or induce
superconductivity at temperatures far above the superconducting critical
temperature (section \ref{photoinduced_unconventional}). In addition, we
should also mention the recent efforts in using intense THz pulses to probe
collective modes of superconductors (for classification of collective modes
in superconductors see e.g. \cite{Millis2020a}).

The amplitude mode of the superconducting order parameter, describing
variations of the amplitude of the order parameter with the frequency of 2$%
\Delta /h$, commonly referred to as the Higgs mode, is a scalar excitation
of the order parameter. As it does not carry spin or charge, the Higgs mode
does not couple linearly to the electromagnetic field. Only when
superconductivity coexists with a charge density wave (CDW) order, as in
2H-NbSe$_{2}$ \cite{Klein,Littlewood,Sacuto}, has Higgs mode been detected
using linear spectroscopy \cite{Klein,Sacuto}. In this case, the amplitude
mode of the CDW order couples to the Higgs mode via modulation of the
density of states at the Fermi level and thereby perturbing the
superconducting condensate.

There were numerous theoretical reports, suggesting that the excitation of
Higgs mode could be realized by quenching the superconducting order on a
time-scale shorter than the mode frequency \cite{Yuzbashyan,Axt}. These were
followed by experimental studies using intense THz pulses, utilizing the THz
pump - THz probe approach \cite{Matsunaga,ShimanoReview} and the third
harmonic generation \cite{ShimanoReview,Matsunaga2}. These results suggest
the Higgs mode to be observed through its nonlinear coupling to the intense
THz light fields. On the other hand, it has been argued that the third
harmonic generation (THG) can also be attributed to the nonlinear optical
excitation of collective lattice-charge fluctuations \cite{Lara}. It has
been argued, that the THG signal due to density fluctuations should in fact
dominate the response in a BCS superconductor \cite{Cea2}. While the
question remains open \cite{ShimanoReview,Cea2,Murotani,Silaev}, we should mention
yet another approach, where the Higgs mode is observed in the linear optical
conductivity spectrum under injection of supercurrrent \cite%
{Moor,ShimanoCurrent}. Finally, in addition to the Higgs mode,
unconventional superconductors with reduced dimensionality and multi-band
superconductors may host a much richer spectrum of collective excitations 
\cite{Millis2020a,Krull,Eremin}. The first experiments, suggesting the
observation of the collective oscillation of the relative phase between the
order parameters in a two-band superconductor MgB$_{2}$, the Leggett mode,
have just been reported \cite{Benfatto}.

Following the short summary, where we briefly introduced another field of
studies of superconductors that has been made possible by the advancement in
generation of intense electromagnetic pulses - the studies of the collective
modes, we should reiterate some of the open questions and the possible ways
of addressing them.

We have shown that the dynamics of the superconducting state of conventional
BCS superconductors, excited by femtosecond optical pulses, can be extremely
well described by the phenomenological Rothwarf-Taylor model \cite%
{RothwarfTaylor,RT}. Surprisingly, the same can be said about experiments in
the weak perturbation regime in unconventional d-wave superconductors,
despite the fact that they are characterized by the anisotropic gap with
nodes. Both pair-breaking and the superconducting state recovery dynamics
are found to be qualitatively consistent with a boson-mediated pairing in
cuprates. The question about the nature of the pairing "glue" is, however,
still open. Here, systematic studies, by varying the energy of excitation
photons, could provide further clues. Recent studies on the optimally doped
PCCO suggest that pairing can be mediated either by bosons or
antiferromagnetic fluctuations \cite{PCCOBeck}. Thus, doping dependent
studies may be able to distinguish between the two scenarios. 

Moreover, photoexcitation of cuprate superconductors seems to give rise to
filling-up of the gap instead of the suppression of its magnitude \cite%
{Dessau}. Here a major progress in understanding can be expected from
systematic tr-ARPES studies with high energy resolution and large momentum
coverage \cite{Damascelli,Dima,ArpesGedik}.

Another rapidly developing field of research focuses on amplification of
superconductivity. Here, studies in both conventional \cite{Beck13} and
unconventional superconductors \cite%
{Fausti,Wu,Mankowsky,Kaiser,CavalleriRev,Averitt} require further
investigations. In NbN \cite{Beck13}, narrow band THz excitation studies
suggest superconducting state enhancement to compete with pair-breaking, yet
the spectrally resolved study is still lacking. Such studies could provide
further evidence of non-thermal quasiparticle distribution, supporting the
proposed scenario of Eliashberg enhancement of superconductivity \cite%
{Beck13}. Moreover, by tuning the excitation photon energy, detailed
information of the underlying processes could be obtained, and compared to
different theoretical proposals \cite{Scalapino,Scalapino2,Chang78}. As far
as the studies of photoinduced enhancement of superconductivity in
unconventional superconductors is concerned, the main advancement could be
expected by studying optically thin films. This way, one could insure a
homogeneous excitation profile, and dramatically enhance the sensitivity.

Finally, in many of the unconventional superconductors, their ground state
properties can be controlled by doping, pressure or strain. Here, a
combination of pressure-tuning and impulsive excitation should provide a new
playground to generate and investigate emergent (metastable) states,
especially when pressure-tuning the materials into the vicinity of (quantum)
critical points in their phase diagrams.  

\section*{Acknowledgements} The work was supported by the Deutsche Forschungsgemeinschaft (DFG, German Research Foundation) - TRR 173 - 26856537. We thank Viktor Kabanov for valuable discussions.


\begin{thebibliography}{999}
\bibitem{Tinkham} M. Tinkham, \textit{Introduction to superconductivity},
McGraw Hill, 1996, Chapter 11.

\bibitem{Gray} K.E. Gray (ed.), \textit{Nonequilibrium Superconductivity,
Phonons and Kapitza Boundaries}, NATO ASI Series, Plenum, New York (1981).

\bibitem{TinkhamGlover} R.E. Glover \& M. Tinkham, \textit{Physical Review} 
\textbf{104}, 844 (1956).

\bibitem{Burstein} E. Burstein, D. N. Langenberg, and B. N. Taylor, \textit{%
Phys. Rev. Lett.} \textbf{6}, 92 (1961).

\bibitem{Wyatt} A.F.G. Wyatt, V. M. Dmitriev, W.S. Moore, and F.W. Sheard, 
\textit{Phys. Bev. Lett.} \textbf{16}, 1166 (1966)

\bibitem{Dayem} A.H. Dayem and J.J. Wiegand, \textit{Phys. Rev.} \textbf{155}%
, 419 (1967)

\bibitem{Klapwijk} T.M. Klapwijk, J.N. van den Bergh, and J.E. Mooij, 
\textit{J. Low Temp. Phys.} \textbf{26}, 385 (1977)

\bibitem{Clarke} T. Kommers and J. Clarke, \textit{Phys. Rev. Lett.} \textbf{%
38}, 1091 (1977).

\bibitem{Federici} J.F. Federici, B. I. Greene, P. N. Saeta, D. R. Dykaar,
F. Sharifi, and R. C. Dynes, \textit{Phys. Rev}. B\textbf{\ 46}, 11153
(1992).

\bibitem{Han} S.G. Han, Z. V. Vardeny, K. S. Wong, O. G. Symko, and G.
Koren, \textit{Phys. Rev. Lett.} \textbf{65}, 2708 (1990).

\bibitem{Chwalek} J.M. Chwalek, C. Uher, J.F. Whitaker, G.A. Morou and J.A.
Agostinelli, \textit{Appl.Phys.Lett.} \textbf{58}, 980 (1991)

\bibitem{Albrecht} W. Albrecht, Th. Kruse and H. Kurz, \textit{Phys. Rev.
Lett.} \textbf{69}, 1451 (1992).

\bibitem{Stevens} C.J. Stevens, D. Smith, C. Chen, J. F. Ryan, B. Podobnik,
D. Mihailovic, G. A. Wagner, and J. E. Evetts, \textit{Phys. Rev. Lett.} 
\textbf{78}, 2212 (1997).

\bibitem{DemsarYBCO} J. Demsar, B. Podobnik, V. V. Kabanov, Th. Wolf, and D.
Mihailovic, Phys. Rev. Lett. \textbf{82}, 4918 (1999).

\bibitem{kabanov} V.V. Kabanov, J. Demsar, B. Podobnik, and D. Mihailovic, 
\textit{Phys. Rev. B} \textbf{59}, 1497 (1999).

\bibitem{Kaindl} R.A. Kaindl, M. Woerner, T. Elsaesser, D.C. Smith, J.F.
Ryan, G.A. Farnan, M.P. McCurry, D.G. Walmsley, \textit{Science} \textbf{287}%
, 470 (2000).

\bibitem{THzAveritt} R.D. Averitt, G. Rodriguez, A. I. Lobad, J. L. W.
Siders, S. A. Trugman, and A. J. Taylor, \textit{Phys. Rev}.\ B\textbf{\ 63}%
, 140502(R) (2001).

\bibitem{Segre} G.P. Segre, N. Gedik, J. Orenstein, D. A. Bonn, Ruixing
Liang, and W. N. Hardy, \textit{Phys. Rev. Lett.} \textbf{88}, 137001 (2002).

\bibitem{Carr} G.L. Carr, R.P.S.M. Lobo, J. LaVeigne, D.H. Reitze, and D.B.
Tanner, \textit{Phys. Rev. Lett.} \textbf{85}, 3001 (2000).

\bibitem{MgB2} J. Demsar, R. D. Averitt, A. J. Taylor, V. V. Kabanov, W. N.
Kang, H. J. Kim, E. M. Choi, and S. I. Lee, \textit{Phys. Rev. Lett.} 
\textbf{91}, 267002 (2003).

\bibitem{Lobo} R.P.S.M. Lobo, J.D. LaVeigne, D.H. Reitze, D.B. Tanner, Z.H.
Barber, E. Jacques, P. Bosland, M.J. Burns, and G.L. Carr, \textit{Phys. Rev.%
} B\textbf{\ 72}, 024510 (2005).

\bibitem{Beck1} M. Beck, M. Klammer, S. Lang, P. Leiderer, V. V. Kabanov, G.
N. Gol'tsman, and J. Demsar, \textit{Phys. Rev. Lett.} \textbf{107}, 177007
(2011).

\bibitem{Giannetti} C. Giannetti, M. Capone, D. Fausti, M. Fabrizio, F.
Parmigiani \& D.Mihailovic, \textit{Advances in Physics}\textbf{\ 65},
58-238 (2016).

\bibitem{Ginsberg62} D. M. Ginsberg, \textit{Phys. Rev. Lett.} \textbf{8},
204 (1962).

\bibitem{Parker72} W. H. Parker and W. D. Williams, \textit{Phys. Rev. Lett.}
\textbf{29}, 924 (1972).

\bibitem{SchriefferGinsberg} J.R. Schrieffer and D.M. Ginsberg, \textit{%
Phys. Rev. Lett.} \textbf{8}, 207 (1962).

\bibitem{MillerDayem} B.I. Miller, A.H. Dayem, \textit{Phys. Rev. Lett.} 
\textbf{8}, 207 (1967).

\bibitem{Gray69} K.E. Gray, A.R. Long, C. J. Adkins, \textit{Philosophical
Mag.} \textbf{20}, 273 (1969)

\bibitem{RothwarfTaylor} A. Rothwarf, B.N. Taylor, \textit{Phys. Rev. Lett. }%
\textbf{19}, 27 (1967).

\bibitem{Testardi} L.R. Testardi, \textit{Phys. Rev. B }\textbf{4}, 2189
(1971).

\bibitem{OwenScalapino} C.S. Owen and D.J. Scalapino, \textit{Phys. Rev.
Lett. }\textbf{28}, 1559 (1972).

\bibitem{SaiHalasz} G.A. Sai-Halasz, C.C. Chi, A. Denenstein, and D.N.
Langenberg, \textit{Phys. Rev. Lett. }\textbf{33}, 215 (1974).

\bibitem{Parker} W. Parker, \textit{Phys. Rev. B} \textbf{12}, 3667 (1975).

\bibitem{Elesin} V.F. Elesin and Yu.V. Kopaev, \textit{Sov. Phys. Usp.} 
\textbf{24}, 116 (1981).

\bibitem{Klapwijk77} T. M. Klapwijk, J. N. van den Bergh, and J. E, Mooij, 
\textit{J. Low Temp. Phys.} \textbf{26}, 385 (1977).

\bibitem{Schmid77} A. Schmid, \textit{Phys. Rev. Lett.} \textbf{38}, 922
(1977).

\bibitem{Eliashberg70} G.M. Eliashberg, \textit{Pis'ma ZETF }\textbf{11},
186 (1970) [\textit{JETP Lett.} \textbf{11}, 114 (1970)]

\bibitem{Eliashberg73} B.I. Ivlev, S.G. Lisitsyn, and G.M Eliashberg, 
\textit{J. Low Temp. Phys.} \textbf{10}, 449 (1973).

\bibitem{Rousseau} I. Rousseau, MSc Thesis, University of Konstanz (2012).

\bibitem{Schoen} U. Eckern, A. Schmid, M. Schmutz, and G. Schon, \textit{J.
Low Temp. Phys.} \textbf{36}, 643 (1979).

\bibitem{Scalapino} J.J. Chang and D.J. Scalapino, \textit{Phys. Rev}. B%
\textbf{\ 15}, 2651 (1977).

\bibitem{Scalapino2} J.J. Chang and D.J. Scalapino, \textit{J. Low Temp.
Phys.} \textbf{29}, 477 (1977).

\bibitem{Chang78} J.J. Chang and D.J. Scalapino, \textit{J. Low Temp. Phys.} 
\textbf{31}, 1 (1978).

\bibitem{Tredwell75} T. J. Tredwell and E. H. Jacobsen, \textit{Phys. Rev.
Lett. }\textbf{35}, 244 (1975).

\bibitem{Tredwell76} T. J. Tredwell and E. H. Jacobsen, \textit{Phys. Rev. B 
}\textbf{13}, 2931 (1976).

\bibitem{Seligson} D. Seligson and J. Clarke, \textit{Phys. Rev. B }\textbf{%
28}, 6297 (1983).

\bibitem{Cirillo} N.C. Cirillo, Jr., W.L. Clinton, and J. R. Waterman, 
\textit{Phys. Rev. B }\textbf{25}, 5698 (1982).

\bibitem{Pals} J.A. Pals, K. Weiss, P. van Attekum, R.E. Horstman, and J.
Wolter, \textit{Physics Reports} \textbf{89}, 323 (1979).

\bibitem{Mooij} J. E. Mooij, in Nonequilibrium superconductivity, phonons,
and Kapitza boundaries (Springer, 1981), p. 191.

\bibitem{Hilton} D.J. Hilton, \textit{in } in \textit{Optical Techniques for
Solid-State Materials Characterization}, Edited by R.P. Prasankumar and A.J.
Taylor (Francis \& Taylor, New York, 2011).

\bibitem{mercury} J. Demsar, R. Hudej, J. Karpinski, V. V. Kabanov, and D.
Mihailovic, \textit{Phys. Rev. }B\textbf{\ 63}, 054519 (2001).

\bibitem{Gedik} N. Gedik, M. Langner, J. Orenstein, S. Ono, Y. Abe, and Y.
Ando, \textit{Phys. Rev. Lett.} \textbf{95}, 117005 (2005).

\bibitem{Kusar} P. Kusar, V.V. Kabanov, J. Demsar, T. Mertelj, S. Sugai, and
D. Mihailovic, \textit{Phys. Rev. Lett. }\textbf{101}, 227001 (2008).

\bibitem{Giannetti09} C. Giannetti, G. Zgrablic, C. Consani, A. Crepaldi, D.
Nardi, G. Ferrini, G. Dhalenne, A. Revcolevschi, and F. Parmigiani, \textit{%
Phys. Rev.} B\textbf{\ 80}, 235129 (2009).

\bibitem{Mertelj} T. Mertelj, V. V. Kabanov, C. Gadermaier, N. D. Zhigadlo,
S. Katrych, J. Karpinski, and D. Mihailovic, \textit{Phys. Rev. Lett.} 
\textbf{102}, 17002 (2009).

\bibitem{Chia} E.E.M. Chia, D. Talbayev, J-X. Zhu, H.Q. Yuan, T. Park, J.D.
Thompson, C. Panagopoulos, G.F. Chen, J.L. Luo, N.L. Wang, and A.J. Taylor, 
\textit{Phys. Rev. Lett.} \textbf{104}, 027003 (2010).

\bibitem{Mansart} B. Mansart, D. Boschetto, A. Savoia, F. Rullier-Albenque,
F. Bouquet, E. Papalazarou, A. Forget, D. Colson, A. Rousse, and M. Marsi, 
\textit{Phys. Rev. }\textbf{B 82}, 024513 (2010).

\bibitem{GedikPnic} D.H. Torchinsky, G. F. Chen, J. L. Luo, N. L. Wang, and
N. Gedik, \textit{Phys. Rev. Lett.} \textbf{105}, 027005 (2010).

\bibitem{KaindlBiSCO} R.A. Kaindl, M.A. Carnahan, D.S. Chemla, S. Oh, and
J.N. Eckstein, \textit{Phys. Rev. } \textbf{B 72}, 060510 (2005).

\bibitem{Pashkin} A. Pashkin, M. Porer, M. Beyer, K. W. Kim, A. Dubroka, C.
Bernhard, X. Yao, Y. Dagan, R. Hackl, A. Erb, J. Demsar, R. Huber, and A.
Leitenstorfer, \textit{Phys. Rev. Lett.} \textbf{105}, 067001 (2010).

\bibitem{Beyer} M. Beyer, D. St\"{a}dter, M. Beck, H. Sch\"{a}fer, V. V.
Kabanov, G. Logvenov, I. Bozovic, G. Koren, and J. Demsar, \textit{Phys. Rev}%
. B\textbf{\ 83}, 214515 (2011).

\bibitem{PCCOBeck} M. Beck, M. Klammer, I. Rousseau, M. Obergfell, P.
Leiderer, M. Helm, V. V. Kabanov, I. Diamant, A. Rabinowicz, Y. Dagan, and
J. Demsar, \textit{Phys. Rev.} B\textbf{\ 95}, 085106 (2017).

\bibitem{Sindler} M. \v{S}indler, C. Kadlec, P. Ku\v{z}el, K. Ilin, M.
Siegel, and H. Nemec, \textit{Phys. Rev.}\ B \textbf{97}, 054507 (2018).

\bibitem{Perfetti} L. Perfetti, P. A. Loukakos, M. Lisowski, U. Bovensiepen,
H. Eisaki, and M. Wolf, \textit{Phys. Rev. Lett.} \textbf{99}, 197001 (2007).

\bibitem{Uwe} R. Cort\'{e}s, L. Rettig, Y. Yoshida, H. Eisaki, M. Wolf, and
U. Bovensiepen, \textit{Phys. Rev. Lett.} \textbf{107}, 097002 (2011).

\bibitem{Lanzara} C.L. Smallwood, J.P. Hinton, C. Jozwiak, W. Zhang, J.D.
Koralek, H. Eisaki, D-H. Lee, J. Orenstein, A. Lanzara, \textit{Science }%
\textbf{336}, 1137 (2012).

\bibitem{ZXShen} W. Lee, J. J. Lee, E. A. Nowadnick, W. Tabis, S. W. Huang,
V.N. Strocov, E. M. Motoyama, G. Yu, B. Moritz, M. Greven, T. Schmitt, Z. X.
Shen, T. P. Devereaux, \textit{Nature Phys} \textbf{10}, 883--889 (2014).

\bibitem{HeHRArpes} Yu. He, I.M. Vishik, M. Yi, S. Yang, Z. Liu, J.J. Lee,
S. Chen, S.N. Rebec, D. Leuenberger, A. Zong, C. M. Jefferson, R.G. Moore,
P.S. Kirchmann, A.J. Merriam, and Z-X. Shen, \textit{Rev. Sci. Inst.} 
\textbf{87}, 011301 (2016).

\bibitem{Dessau} S. Parham, H. Li, T.\thinspace J. Nummy, J.\thinspace A.
Waugh, X.\thinspace Q. Zhou, J. Griffith, J. Schneeloch, R.\thinspace D.
Zhong, G.\thinspace D. Gu, and D.\thinspace S. Dessau, \textit{Phys. Rev. X} 
\textbf{7}, 041013 (2017).

\bibitem{Bauer} I. Avigo et al., \textit{Phys. Status Solidi }B \textbf{254}%
, 1600382 (2017).

\bibitem{Emitter} M. Beck, H. Sch\"{a}fer, G. Klatt, J. Demsar, S. Winnerl,
M. Helm, and T. Dekorsy, \textit{Opt. Express} \textbf{18}, 9251 (2010) .

\bibitem{Zimmermann} W. Zimmermann, E.H. Brandt, M.Bauer, E.Seider,
L.Genzel, \textit{Physica} \textbf{C} \textbf{183}, 99 (1991).

\bibitem{AllenEx} P.B. Allen, \textit{Phys. Rev. Lett. }\textbf{59}, 1460
(1987).

\bibitem{BrorsonEx} S.D. Brorson, A. Kazeroonian, J.S. Moodera, D.W. Face,
T.K. Cheng, E.P. Ippen, M.S. Dresselhaus, and G. Dresselhaus, \textit{Phys.
Rev. Lett. }\textbf{64}, 2172 (1990).

\bibitem{BookChapter} J. Demsar und T. Dekorsy, in \textit{Optical
Techniques for Solid-State Materials Characterization}, Edited by R.P.
Prasankumar and A.J. Taylor (Francis \& Taylor, New York, 2011).

\bibitem{KabanovTTM} V.V. Baranov and V.V. Kabanov, \textit{Phys. Rev}. B%
\textbf{\ 89}, 125102 (2014).

\bibitem{Obergfell} M. Obergfell, J. Demsar, \textit{Phys. Rev. Lett. }%
\textbf{124}, 037401 (2020).

\bibitem{escape} A. Rothwarf, G.A. Sai-Halasz, and D.N. Langensberg, \textit{%
Phys. Rev. Lett. }\textbf{33}, 212, (1974).

\bibitem{RT} V.V. Kabanov, J. Demsar, D. Mihailovic, \textit{Phys. Rev. Lett.%
} \textbf{95}, 147002 (2005).

\bibitem{Ovchinikov} Yu. N. Ovchinnikov and V.Z. Kresin, \textit{Phys. Rev}.
B\textbf{\ 58}, 12416 (1998).

\bibitem{Geibel} C. Geibel, H. Rietschel, A. Junod, M. Pelizzone and J.
Muller, \textit{J. Phys. F: Met. Phys. }\textbf{15}, 405 (1985).

\bibitem{Weber} W. Weber, \textit{Phys. Rev}. B\textbf{\ 8}, 5082 (1973).

\bibitem{lambda} U. Haufe, G. Kerker and K.H. Bennemann, \textit{Sol. Stat.
Comm.} \textbf{17}, 321 (1975).

\bibitem{allen} P.B. Allen, \textit{Phys. Rev. Lett. }\textbf{59}, 1460
(1987).

\bibitem{GedikYBCO} N. Gedik, P. Blake, R.C. Spitzer, J. Orenstein, R.
Liang, D.A. Bonn, and W. Hardy, Phys. Rev. B \textbf{70}, 014504, (2004).

\bibitem{Hinton} J. P. Hinton, E. Thewalt, Z. Alpichshev, F. Mahmood, J. D.
Koralek, M. K. Chan, M. J. Veit, C. J. Dorow, N. Bari\v{s}i\'{c}, A. F.
Kemper, D. A. Bonn, W. N. Hardy, Ruixing Liang, N. Gedik, M. Greven, A.
Lanzara \& J. Orenstein, \textit{Scientific Reports} \textbf{6}, 23610
(2016).

\bibitem{Orenstein} J. Orenstein, \textit{Phys. Today} \textbf{65}, 44
(2012).

\bibitem{Sobolev} S. Sobolev, T. Dong, N. Bhattacharjee, A. Lanz, A.R.
Pokharel, et al., in preparation

\bibitem{Ilin} K.S. Il'in, M. Lindgren, M. Currie, A. D. Semenov, G. N.
Gol'tsman, and Roman Sobolewski, \textit{Appl. Phys. Lett.} \textbf{76},
2752 (2000).

\bibitem{HFprl} J. Demsar, V.K. Thorsmolle, J.L. Sarrao, A.J. Taylor, 
\textit{Phys. Rev. Lett}. \textbf{96}, 037401 (2006).

\bibitem{HFreview} J. Demsar, J.L. Sarrao, A.J.Taylor, \ \textit{J. Phys.:
Cond. Mat.} \textbf{18}, R281 (2006).

\bibitem{Ahn} K. H. Ahn, M. J. Graf, S. A. Trugman, J. Demsar, R. D.
Averitt, J. L. Sarrao, and A. J. Taylor,\textit{\ Phys. Rev. }\textbf{B} 69,
045114 (2004).

\bibitem{HFPaper} J. Demsar, R.D. Averitt, K.H. Ahn, M.J. Graf, S.A.
Trugman, V.V. Kabanov, J.L. Sarrao, A.J. Taylor, \textit{Phys. Rev. Lett.} 
\textbf{91}, 027401 (2003).

\bibitem{Rick} K. S. Burch, E.E.M. Chia, D. Talbayev, B.C. Sales, D.
Mandrus, A. J. Taylor, and R. D. Averitt, \textit{Phys. Rev. Lett.} \textbf{%
100}, 026409 (2008).

\bibitem{YbAl} J. Demsar, V.V. Kabanov, A.S. Alexandrov, H.J. Lee, E.D.
Bauer, J.L. Sarrao, A.J. Taylor, \textit{Phys. Rev. }\textbf{B} 80, 085121
(2009).

\bibitem{Toni} E.E.M. Chia, Jian-Xin Zhu, D. Talbayev, H. J. Lee, Namjung
Hur, N. O. Moreno, R. D. Averitt, J. L. Sarrao, and A. J. Taylor, \textit{%
Phys. Rev. }\textbf{B} \textbf{84}, 174412 (2011).

\bibitem{DMM1998} D. Mihailovic, B. Podobnik, J. Demsar, G. Wagner, and J.
Evetts, \textit{J. Phys. Chem. Solids }\textbf{59}, 1937 (1998).

\bibitem{Schuller} I. Schuller and K. E. Gray,\textit{\ Phys. Rev. Lett. }%
\textbf{36}, 429 (1976).

\bibitem{SchmidSchoen} A. Schmid and G. Schoen, \textit{J. Low Temp. Phys.} 
\textbf{20}, 207 (1975).

\bibitem{Dvorsek} D. Dvorsek, et al., \textit{Phys. Rev.B} \textbf{66},
020510 (2002).

\bibitem{KusarLSCO} P. Kusar, J. Demsar, D. Mihailovic, and S. Sugai, 
\textit{Phys. Rev. B} \textbf{72} 014544 (2005).

\bibitem{Liu} Y. H. Liu, Y. Toda, K. Shimatake, N. Momono, M. Oda, and M.
Ido, \textit{Phys. Rev. Lett.} \textbf{101}, 137003 (2008).

\bibitem{Luo} C.W. Luo, H. P. Lo, C. H. Su, I. H. Wu, Y.-J. Chen, K. H. Wu,
J.-Y. Lin, T. M. Uen, J. Y. Juang, and T. Kobayashi, \textit{Phys. Rev. B} 
\textbf{82} 104512 (2010).

\bibitem{TodaBisco} Y. Toda, F. Kawanokami, T. Kurosawa, M. Oda, I. Madan,
T. Mertelj, V.V. Kabanov, D. Mihailovic, \textit{Phys. Rev}. B\textbf{\ 90},
094513 (2014).

\bibitem{Kumar} S. Kumar, L. Harnagea, S. Wurmehl, B. Buechner, and A. K.
Sood, EPL 105, 47004 (2014).

\bibitem{Naito} T. Naito, Y. Yamada, T. Inabe, and Y. Toda, \textit{J. Phys.
Soc. Jpn. }\textbf{77}, 064709 (2008).

\bibitem{Zagar} D. Mihailovic, V.V. Kabanov, K. \v{Z}agar, and J. Demsar, 
\textit{Phys. Rev. B} \textbf{60}, R6995(R), (1999).

\bibitem{Schneider} M.L. Schneider, et al., \textit{Europhys. Lett. }\textbf{%
60}, 460 (2002).

\bibitem{GedikBiSCOdoping} N. Gedik, M. Langner, J. Orenstein, S. Ono, Y.
Abe, Y. Ando, \textit{Phys. Rev. Lett.} \textbf{95}, 117005 (2005).

\bibitem{nicol} E. J. Nicol and J. P. Carbotte, \textit{Phys. Rev. }B\textbf{%
\ 67}, 214506 (2003).

\bibitem{Hirschfeld} P. C. Howell, A. Rosch, P. J. Hirschfeld, \textit{Phys.
Rev. Lett. }\textbf{92}, 037003 (2004).

\bibitem{CpLSCO} T. Matsuzaki, N. Momono, M. Oda, and M. Ido, \textit{J.
Phys. Soc. Jpn.} \textbf{73}, 2232 (2004).

\bibitem{Beck13} M. Beck, I. Rousseau, M. Klammer, P. Leiderer, M.
Mittendorff, S. Winnerl, M. Helm, G. N. Gol'tsman, and J. Demsar, \textit{%
Phys. Rev. Lett.} \textbf{110}, 267003 (2013).

\bibitem{Fausti} D. Fausti, R.I. Tobey, N. Dean, S. Kaiser, A. Dienst, M.C.
Hoffmann, S. Pyon, T. Takayama, H. Takagi, A. Cavalleri, \textit{Science }%
\textbf{331}, 189 (2011).

\bibitem{Wu} W. Hu, S. Kaiser, D. Nicoletti, C. R. Hunt, I. Gierz, M. C.
Hoffmann, M. Le Tacon, T. Loew, B. Keimer and A. Cavalleri, \textit{Nat.
Mater.} \textbf{13}, 705 (2014).

\bibitem{Mankowsky} R. Mankowsky, A. Subedi, M. F\"{o}rst, S.O. Mariager, M.
Chollet, H.T. Lemke, J.S. Robinson, J.M. Glownia, M.P. Minitti, A. Frano, M.
Fechner, N.A. Spaldin, T. Loew, B. Keimer, A. Georges \& A. Cavalleri, 
\textit{Nature} \textbf{516}, 71 (2014).

\bibitem{Kaiser} S. Kaiser, C. R. Hunt, D. Nicoletti, W. Hu, I. Gierz, H. Y.
Liu, M. Le Tacon, T. Loew, D. Haug, B. Keimer, and A. Cavalleri, \textit{%
Phys. Rev. B }\textbf{89}, 184516 (2014).

\bibitem{Mitrano} M. Mitrano, A. Cantaluppi, D. Nicoletti, S. Kaiser, A.
Perucchi, S. Lupi, P. Di Pietro, D. Pontiroli, M. Ricco, S. R. Clark, D.
Jaksch and A. Cavalleri, \textit{Nature} \textbf{530}, 461 (2016).

\bibitem{NatPhys} J. Demsar, \textit{Nature Physics} \textbf{12}, 202 (2016).

\bibitem{Cantaluppi} A. Cantaluppi, M. Buzzi, G. Jotzu, D. Nicoletti, M.
Mitrano, D. Pontiroli, M. Ricc\`{o}, A. Perucchi, P. Di Pietro \& A.
Cavalleri, \textit{Nature Physics} \textbf{14}, 837 (2018).

\bibitem{Budden} M. Budden, T. Gebert, M. Buzzi, G. Jotzu, E. Wang, T.
Matsuyama, G. Meier, Y. Laplace, D. Pontiroli, M. Ricc\`{o}, F. Schlawin, D.
Jaksch, A. Cavalleri, \textit{arXiv:}2002.12835

\bibitem{Organics} M. Buzzi, D. Nicoletti, M. Fechner, N. Tancogne-Dejean,
M. A. Sentef, A. Georges, M. Dressel, A. Henderson, T. Siegrist, J. A.
Schlueter, K. Miyagawa, K. Kanoda, M.-S. Nam, A. Ardavan, J. Coulthard, J.
Tindall, F. Schlawin, D. Jaksch, A. Cavalleri, \textit{arXiv:}2001.05389

\bibitem{Foerst} M. F\"{o}rst, C. Manzoni, S. Kaiser, Y. Tomioka, Y. Tokura,
R. Merlin \& A. Cavalleri, \textit{Nature Physics} \textbf{7}, 854 (2011).

\bibitem{CavalleriRev} A. Cavalleri, \textit{Contemporary Physics} \textbf{59%
}, 31-46 (2018).

\bibitem{Averitt} K.A. Cremin, J. Zhang, C.C. Homes, G.D. Guc, Z. Suna, M.M.
Foglera, A.J. Millis, D.N. Basov, and R.D. Averitt, \textit{Proc. Natl.
Acad. Sci.} \textbf{116}, 19875--19879 (2019).

\bibitem{Dodge} J. Orenstein and J. S. Dodge, \textit{Phys. Rev. B} \textbf{%
92}, 134507 (2015).

\bibitem{Response} D. Nicoletti, M. Mitrano, A. Cantaluppi, A. Cavalleri,%
\textit{\ arXiv:1506.07846}

\bibitem{Shimano} H. Niwa, N. Yoshikawa, K. Tomari, R. Matsunaga, D. Song,
H. Eisaki, and R. Shimano, \textit{Phys. Rev. B} \textbf{100}, 104507 (2019)

\bibitem{NLWang} S.J. Zhang, Z.X. Wang, H. Xiang, X. Yao, Q.M. Liu, L.Y.
Shi, T. Lin, T. Dong, D. Wu, and N.L. Wang, \textit{Phys. Rev.} X \textbf{10}%
, 011056 (2020).

\bibitem{Demler} M. Babadi, M. Knap, I. Martin, G. Refael, and E. Demler, 
\textit{Phys. Rev. B }\textbf{96}, 014512 (2017).

\bibitem{Manske} N. Bittner, T. Tohyama, S. Kaiser, and D. Manske, \textit{%
J. Phys. Soc. Jpn.} \textbf{88}, 044704 (2019).

\bibitem{Nava} A. Nava, C. Giannetti, A. Georges, E. Tosatti, and M.
Fabrizio, \textit{Nat. Phys.} \textbf{14}, 154 (2018).

\bibitem{Millis} G. Chiriac\`{o}, A.J. Millis, and I.L. Aleiner, \textit{%
Phys. Rev. B} \textbf{98}, 220510(R) (2018).

\bibitem{Millis2020} G. Chiriac\`{o}, A.J. Millis, and I.L. Aleiner, \textit{%
Phys. Rev. B }\textbf{101}, 041105(R) (2020).

\bibitem{Wang2018} S.J. Zhang, Z.X. Wang, L.Y. Shi, T. Lin, M.Y. Zhang, G.D.
Gu, T. Dong, and N.L. Wang, \textit{Phys. Rev. B }\textbf{98}, 020506(R)
(2018).

\bibitem{Wang2018a} S.J. Zhang, Z.X. Wang, D. Wu, Q.M. Liu, L.Y. Shi, T.
Lin, S.L. Li, P.C. Dai, T. Dong, and N.L. Wang, \textit{Phys. Rev. B }%
\textbf{98}, 224507 (2018).

\bibitem{Liu2019} B. Liu, M. F\"{o}rst, M. Fechner, D. Nicoletti, J. Porras,
B. Keimer, A. Cavalleri, \textit{arXiv:}1905.08356

\bibitem{Kennes} D.M. Kennes, E.Y. Wilner, D.R. Reichman, A.J. Millis, 
\textit{Nat. Phys.} \textbf{13}, 479 (2017).

\bibitem{Kim17} M. Kim, Y. Nomura, M. Ferrero, P. Seth, O. Parcollet, and A.
Georges, \textit{Phys. Rev. B }\textbf{94}, 155152 (2016).

\bibitem{Mazza} G. Mazza, and A. Georges, \textit{Phys. Rev. B} \textbf{96},
064515 (2017).

\bibitem{CpK3C60} K. Allen and F. Hellman, \textit{Phys. Rev. B} \textbf{60}%
, 11765 (1999).

\bibitem{Millis2020a} Z. Sun, M.M. Fogler, D.N. Basov, and A.J. Millis, 
\textit{arXiv:}2001.03704v1

\bibitem{Klein} R. Sooryakumar and M. V. Klein, \textit{Phys. Rev. Lett}. 
\textbf{45}, 660 (1980).

\bibitem{Littlewood} P. B. Littlewood and C. M. Varma, \textit{Phys. Rev.
Lett.} \textbf{47}, 811 (1981).

\bibitem{Sacuto} M.-A. M\'{e}asson, Y. Gallais, M. Cazayous, B. Clair, P.
Rodi\`{e}re, L. Cario, and A. Sacuto, \textit{Phys. Rev. B} \textbf{89},
060503(R) (2014).

\bibitem{Yuzbashyan} E.A. Yuzbashyan, O. Tsyplyatyev, B.L. Altshuler, 
\textit{Phys. Rev. Lett.} \textbf{96}, 097005 (2006).

\bibitem{Axt} T. Papenkort, V.M. Axt, and T. Kuhn, \textit{Phys. Rev. B} 
\textbf{76}, 224522 (2007).

\bibitem{Matsunaga} R. Matsunaga, Y.I. Hamada, K. Makise, Y. Uzawa, H.
Terai, Z. Wang, and R. Shimano, \textit{Phys. Rev. Lett. }\textbf{111},
057002 (2013).

\bibitem{ShimanoReview} R. Shimano and N. Tsuji, \textit{Annu. Rev. Condens.
Matter Phys.} \textbf{11}, 103--24 (2020).

\bibitem{Matsunaga2} R. Matsunaga, N. Tsuji, H. Fujita, A. Sugioka, K.
Makise, Y. Uzawa, H. Terai, Z. Wang, H. Aoki, and R. Shimano, \textit{Science%
} \textbf{345}, 1145 (2014).

\bibitem{Lara} T. Cea, C. Castellani, and L. Benfatto, \textit{Phys. Rev. B} 
\textbf{93}, 180507(R) (2016).

\bibitem{Cea2} T. Cea, P. Barone, C. Castellani, and L. Benfatto, \textit{%
Phys. Rev. B} \textbf{97}, 094516 (2018).

\bibitem{Murotani} Y. Murotani, R. Shimano, \textit{Phys. Rev. B} \textbf{99}%
, 224510 (2019)

\bibitem{Silaev} M. Silaev, \textit{Phys. Rev. B} \textbf{99}, 224511 (2019).


\bibitem{Moor} A. Moor, A. F. Volkov, and K. B. Efetov, \textit{Phys. Rev.
Lett.} \textbf{118}, 047001 (2017).

\bibitem{ShimanoCurrent} S. Nakamura, Y. Iida, Y. Murotani, R. Matsunaga, H.
Terai, and R. Shimano, \textit{Phys. Rev. Lett. }\textbf{122}, 257001 (2019).

\bibitem{Krull} H. Krull, N. Bittner, G.S. Uhrig, D. Manske and A. P.
Schnyder, \textit{Nature Communications} \textbf{7}, 11921 (2016).

\bibitem{Eremin} M.A. M\"{u}ller, P.A. Volkov, I. Paul, and I.M. Eremin, 
\textit{Phys. Rev. B} \textbf{100}, 140501(R) (2019).

\bibitem{Benfatto} F. Giorgianni, T. Cea, C. Vicario, C.P. Hauri, W.K.
Withanage, X. Xi and L. Benfatto, \textit{Nature Physics} \textbf{15}, 341
(2019).

\bibitem{Damascelli} A.K. Mills, S. Zhdanovich, M.X. Na, F. Boschini, E.
Razzoli, M. Michiardi, A. Sheyerman, M. Schneider, T.J. Hammond, V. S\"{u}%
ss, C. Felser, A. Damascelli, D,J. Jones, \textit{Rev. Sci. Inst.} \textbf{90%
}, 083001 (2019).

\bibitem{Dima} D. Kutnyakhov, et al., \textit{Rev. Sci. Inst.} \textbf{91},
013109 (2020).

\bibitem{ArpesGedik} C. Lee, T. Rohwer, E.J. Sie, A. Zong, E. Baldini, J.
Straquadine, P. Walmsley, D. Gardner, Y.S. Lee, I.R. Fisher, N. Gedik, 
\textit{arXiv:1910.14068}
\end{thebibliography}
\end{document}